\documentclass[
 reprint,
 amsmath,amssymb,
 aps,prd
]{revtex4-2}

\usepackage{graphicx}
\usepackage{dcolumn}
\usepackage{bm}
\usepackage{physics}
\usepackage{simplewick}
\usepackage[dvipsnames]{xcolor}
\usepackage{braket,cancel}
\definecolor{lgray}{gray}{0.4}
\definecolor{cor}{rgb}{0.94, 0.41, 0.35}
\definecolor{mag}{rgb}{0.7451, 0.2039, 0.3333}
\usepackage[colorlinks=true,urlcolor=mag,linkcolor=mag,
citecolor=mag,pdfpagelabels=true,hypertexnames=true,
plainpages=false,naturalnames=false]{hyperref}
\newcommand{\x}{{\boldsymbol x}}

\def\k{{\boldsymbol k}}

\def\be{\begin{equation}}
\def\ee{\end{equation}}
\def\ba{\begin{array}}
\def\ea{\end{array}}
\def\nn{\nonumber}

\usepackage{tikz}

\newcommand{\SKSIZE}{9pt}
\newcommand{\SKLINE}{0.8pt}
\newcommand{\SKBOXSIZE}{7pt}

\tikzset{
  sk/bulk+/.style = {
    circle,
    draw=black,
    fill=black,
    minimum size=\SKSIZE,
    inner sep=0pt
  },
  sk/bulk-/.style = {
    circle,
    draw=black,
    fill=white,
    minimum size=\SKSIZE,
    inner sep=0pt
  },
  sk/boundary/.style = {
    rectangle,
    draw=black,
    fill=white,
    minimum width=\SKBOXSIZE,
    minimum height=\SKBOXSIZE,
    inner sep=0pt
  },
  sk/prop/.style = {
    line width=\SKLINE,
    line cap=round
  }
}

\newcommand{\skbulkplus}[2]{%
  \node[sk/bulk+] (#1) at #2 {};%
}

\newcommand{\skbulkminus}[2]{%
  \node[sk/bulk-] (#1) at #2 {};%
}

\newcommand{\skboundary}[2]{%
  \node[sk/boundary] (#1) at #2 {};%
}

\newcommand{\skprop}[2]{%
  \draw[sk/prop] (#1) -- (#2);%
}

\begin{document}

\title{Pushing the Primordial Frontier:
Cosmological Collider Signatures at Strong Mixing}

\author{Javier Huenupi$^a$}

\author{Claudio Mu\~noz$^b$}

\author{Gonzalo A. Palma$^c$}

\author{Spyros Sypsas$^{d}$}

\affiliation{
$^a${\it Center for Particle Cosmology, Department of Physics and Astronomy, University of Pennsylvania, Philadelphia, PA 19104, USA
}
\\
$^b${\it Departamento de Ingenier\'ia Matem\'atica and Centro de Modelamiento Matem\'atico (IRL
2807 CNRS), Universidad de Chile, Casilla 170 Correo 3, Santiago, Chile}
\\
$^c${\it Departamento de F\'isica, FCFM, Universidad de Chile, Blanco Encalada 2008, Santiago, Chile}
\\
$^d${\it Centro de Ciencias Exactas, Facultad de Ciencias, Universidad del B\'io-B\'io, Chill\'an, Chile}
}

\begin{abstract}
We develop an analytic treatment of primordial non-Gaussianity in multifield inflation that is nonperturbative in the constant curvature--isocurvature mixing strength $\lambda$. Using exact linear solutions for the coupled curvature perturbation $\zeta$ and isocurvature perturbation $\sigma$, we construct dressed propagators and derive exact integral representations for the tree-level bispectra generated by the interactions $\dot\zeta^2\sigma$, $\dot\zeta\,\sigma^2$, and $\sigma^3$. This formalism resums curvature--isocurvature transfer to all orders in $\lambda$ and applies for arbitrary values of the entropy mass $\mu$. We obtain closed-form expressions for the leading squeezed limit of the bispectrum contributions and recover the well-known cosmological-collider and quasi-single-field results in the weak-mixing limit. In the strong-mixing regime, where conventional transfer perturbation theory breaks down, we find that both the power spectrum and the bispectrum can be dramatically enhanced, giving rise to distinctive nonperturbative scaling laws for the reduced non-Gaussian amplitude. Together, these results open a new analytic window onto multifield inflation beyond the weak-coupling approximation and establish a framework for studying cosmological-collider signals, primordial-black-hole production, and loop corrections in strongly mixed inflationary dynamics.
\end{abstract}

                          
\maketitle


\section{Introduction}
\label{sec:Intro}

One of the central goals of primordial cosmology is to reconstruct the statistics of $\zeta$, the curvature perturbation that seeded the large-scale structure of our universe. Such a reconstruction may reveal the symmetries, particle content, and interactions that governed cosmic inflation~\cite{Guth:1980zm,Starobinsky:1980te,Linde:1981mu,Albrecht:1982wi,Achucarro:2022qrl}. Primordial non-Gaussianity is especially informative in this regard~\cite{Bartolo:2004if,Chen:2010xka,Wang:2013zva}, as it carries information beyond the power spectrum and directly probes the interactions of the early universe. Although the full probability distribution of $\zeta$ generally lies beyond conventional perturbative methods~\cite{Chen:2018brw,Chen:2018uul}, much of its physical content remains accessible through low-order correlation functions, computed using the in-in formalism~\cite{Weinberg:2005vy,Adshead:2009cb} or Schwinger--Keldysh diagrammatics~\cite{Adshead:2009cb,Chen:2017ryl}. The first of these directly sensitive to primordial non-Gaussianity is the three-point function, or bispectrum~\cite{Gangui:1993tt,Acquaviva:2002ud,Maldacena:2002vr,Liguori:2010hx}. Its dependence on the three external momenta---its shape in Fourier space---encodes not only the strength of primordial interactions, but also the dynamical mechanisms and characteristic time scales that shaped the primordial fluctuations.

In canonical single-field slow-roll inflation, the interactions responsible for the bispectrum are themselves suppressed by slow-roll parameters. More general single-field theories can instead admit sizeable derivative interactions, systematically organized within the effective field theory of inflation~\cite{Cheung:2007st}. These interactions imprint characteristic shapes on the momentum dependence of the bispectrum~\cite{Babich:2004gb}, most notably the equilateral and orthogonal templates. Their amplitudes, conventionally parametrized by the nonlinearity parameter $f_{\rm NL}$, are constrained by Planck~\cite{Planck:2019kim} as $f_{\rm NL}^{\rm equil} = -26 \pm 47$, and $f_{\rm NL}^{\rm ortho} = -38 \pm 24$. Within concrete models, these bounds constrain underlying quantities such as the sound speed of primordial fluctuations~\cite{Planck:2019kim,Senatore:2009gt}, thereby providing a window onto possible ultraviolet completions.

Beyond probing the self-interactions of the inflaton, the bispectrum is sensitive to additional degrees of freedom active during inflation. Such fields arise naturally in ultraviolet-complete constructions, where inflation typically unfolds on a multidimensional field space~\cite{Baumann:2014nda}. A turn of the background trajectory induces mixing between the curvature perturbation and isocurvature fluctuations orthogonal to the background motion. Their subsequent conversion into curvature perturbations can source local non-Gaussianity~\cite{Komatsu:2001rj,Lyth:2005fi,Lyth:2002my}, whose amplitude is constrained by Planck as $f_{\rm NL}^{\rm local}=-0.9\pm5.1$~\cite{Planck:2019kim}. More distinctively, additional fields can imprint nonanalytic features on the squeezed limit of the bispectrum. Such signatures lie at the heart of quasi-single-field inflation and the cosmological-collider program~\cite{Chen:2009zp, Chen:2009we, Chen:2012ge, Noumi:2012vr,Arkani-Hamed:2015bza, Lee:2016vti, Meerburg:2016zdz, Chen:2016uwp, An:2017hlx, Iyer:2017qzw, Wang:2019gok, Aoki:2020zbj, Pinol:2021aun, Chakraborty:2023qbp, Cabass:2024wob, Xianyu:2025lbk, Kumar:2026ogn, Kumar:2026dih, Philcox:2026rpn}, which seek to use primordial correlators as probes of particles with masses of order the Hubble scale.

In this work, we consider the simplest setting in which these effects arise: a two-field inflationary model with a turning background trajectory. To leading order in slow roll, the quadratic action for the curvature perturbation $\zeta$ and an isocurvature perturbation $\sigma$ is
\begin{align}
S^{(2)}
&= \!\!
\int_{x} \! a^3
\Big[
m_{\rm Pl}^2 \epsilon \Big( 
\left(D_t \zeta
\right)^2  - \frac{1}{a^2}(\nabla\zeta)^2 \Big)  \nn\\
& \qquad
+\frac12\dot\sigma^2
-\frac{1}{2a^2}(\nabla\sigma)^2
-\frac12\mu^2\sigma^2
\Big],
\label{S2}
\end{align}
where $\int_{x} \equiv \int \dd^4 x$. Here $m_{\rm Pl}$ is the reduced Planck mass, $a(t)$ is the scale factor, $H=\dot a/a$ is the Hubble parameter, $\epsilon=-\dot H/H^2$ is the first slow-roll parameter and $\mu$ is the entropy mass. The kinetic term for $\zeta$ includes a covariant time derivative $D_t \zeta$ defined as
\be \label{cov-D-t}
D_t \zeta \equiv \dot\zeta
-\frac{\lambda}{\sqrt{2\epsilon}}\frac{H}{m_{\rm Pl}}\,\sigma ,
\ee
where dimensionless parameter $\lambda$ controls the linear mixing between the curvature and isocurvature perturbations~\cite{Gordon:2000hv,GrootNibbelink:2001qt,Langlois:2008qf,Tolley:2009fg}. The phenomenology of this system depends nontrivially on both $\mu$ and $\lambda$. In particular, the transfer between the curvature and isocurvature sectors can modify or enhance the power spectrum, with applications ranging from CMB observables~\cite{Achucarro:2010da,Achucarro:2012fd,Gao:2013ota,Achucarro:2013cva,Mizuno:2014tza,Achucarro:2014msa,Gao:2015aba,Noumi:2013cfa,Pi:2012gf} to PBH formation~\cite{Garcia-Bellido:1996mdl,Palma:2020ejf,Fumagalli:2020adf,Braglia:2020eai,Pi:2021dft,Anguelova:2020nzl,Geller:2022nkr,Lorenzoni:2025gni,Boutivas:2022qtl}. However, most analytic studies of higher-order correlators~\cite{Seery:2005gb,Rigopoulos:2005ae,Vernizzi:2006ve,Byrnes:2008wi,GarciaSaenz:2018vqf,Fumagalli:2019noh,Bjorkmo:2019aev,Garcia-Saenz:2019njm,Iarygina:2023rapid,Aoki:2026qea,Lalak:2007vi,Konieczka:2014zja,Achucarro:2019lgo,Fumagalli:2020nvq,Firouzjahi:2021lov,Achucarro:2016fby,Achucarro:2026qyx,Kaiser:2013sna,Mizuno:2017idt,Achucarro:2018ngj,Nguyen:2019kbm,RenauxPetel:2015mga,GarciaSaenz:2018ifx,Aragam:2019omo,Aragam:2020uqi,Christodoulidis:2023eiw,Wolters:2024rapid,Anguelova:2024akm} treat the quadratic mixing as a small interaction and expand perturbatively in $\lambda$.

In Ref.~\cite{Huenupi:2026abj}, we obtained exact analytic solutions to the linear equations derived from Eq.~\eqref{S2}, valid for arbitrary entropy mass $\mu$ and constant mixing strength $\lambda$. As a first application, we used these solutions to express the primordial power spectrum in closed form. The present work develops this formalism in detail and extends it to the bispectrum. By incorporating the quadratic mixing directly into the exact mode functions and propagators, rather than treating it perturbatively as an interaction, we resum curvature--isocurvature transfer to all orders in $\lambda$. To this end, we supplement the quadratic action with three representative cubic interactions
\be 
S_{\rm int}^{(3)} \!
=  \!\!
\int_{x} \! a^3   \!
\left[ 
\frac{\widetilde{\alpha} \sqrt{ \epsilon} m_{\rm Pl}}{\sqrt{2} }\,\sigma\dot\zeta^2
+  \frac{\widetilde{\beta} H}{2}\,
\sigma^2\dot\zeta
+ \frac{\widetilde{\gamma}H^2}{3! m_{\rm Pl} \sqrt{2 \epsilon}}\sigma^3
\right] \!,
\label{S3}
\ee
where $\widetilde{\alpha}$, $\widetilde{\beta}$, and $\widetilde{\gamma}$ are dimensionless constant cubic couplings. Using the exact linear solutions, we construct dressed Schwinger--Keldysh propagators and derive exact integral representations for the tree-level bispectra generated by each interaction in Eq.~\eqref{S3}. We then extract their behavior in the squeezed configuration $k_L\ll k_S$, in which one external wavelength is much longer than the other two. For a light but massive entropy field, $0<\mu/H<3/2$, the leading nonanalytic contributions to the squeezed bispectra take the closed form
\be
\mathcal S(k_S,k_S,k_L)
\propto
\sum_{\pm}
\mathcal C_{\pm}(\lambda,\nu)
\left(\frac{k_L}{k_S}\right)^{\frac12\pm\nu},
\ee
where $\nu\equiv\sqrt{9/4-\mu^2/H^2}$ is determined entirely by the entropy mass, while the coefficients $\mathcal C_{\pm}(\lambda,\nu)$ depend nonperturbatively on the mixing strength. Analytic continuation to the heavy regime, $\nu=i\rho$, with $\rho\equiv\sqrt{\mu^2/H^2-9/4}$, yields the logarithmic oscillations characteristic of heavy-particle exchange,
\be
\mathcal S(k_S,k_S,k_L)
\propto
\left(\frac{k_L}{k_S}\right)^{1/2}
\mathcal C(\lambda,\rho)
\sin\left[
\rho\ln\left(\frac{k_S}{k_L}\right)+\phi
\right].
\ee
In the weak-mixing limit, our expressions reproduce the familiar quasi-single-field and cosmological-collider results~\cite{Chen:2009zp,Meerburg:2016zdz,Lee:2016vti}. In the strong-mixing regime, they reveal a simultaneous and potentially dramatic enhancement of the power spectrum and bispectrum, together with distinctive nonperturbative scaling laws for the reduced non-Gaussian amplitude.

A particularly striking aspect of these results is the precise dependence of the squeezed-limit shape $\mathcal S(k_S,k_S,k_L)$ on both $\lambda$ and $\mu$. In a perturbative treatment of the mixing, the quadratic action in Eq.~\eqref{S2} is conventionally decomposed as~\cite{Cespedes:2012hu,Achucarro:2012sm,Achucarro:2012yr,An:2017hlx,Iyer:2017qzw}
\begin{align}
S^{(2)}
&=
\int_x \! a^3
\Bigg[
m_{\rm Pl}^2 \epsilon\dot\zeta^2
- m_{\rm Pl}^2 \frac{\epsilon}{a^2}(\nabla\zeta)^2
\nn\\
& \quad
+\frac12\dot\sigma^2
-\frac{1}{2a^2}(\nabla\sigma)^2
-\frac12m_{\rm eff}^2\sigma^2
\Bigg]
+S^{(2)}_{\rm int},
\end{align}
where $m_{\rm eff}^2\equiv\mu^2-H^2\lambda^2$ is the so-called effective mass, while $S^{(2)}_{\rm int}$ contains the quadratic mixing:
\be
S^{(2)}_{\rm int}
=
- m_{\rm Pl} \int_x \! a^3\lambda \sqrt{2\epsilon} H\dot\zeta\,\sigma.
\ee
Our results show that, in the strong-mixing regime, the oscillatory behavior of the squeezed bispectrum is governed not by $m_{\rm eff}^2$, but by the combination $\mu^2=m_{\rm eff}^2 + \lambda^2 H^2$, in agreement with previous nonperturbative analyses of quasi-single-field inflation~\cite{An:2017hlx}.

The exact treatment of the quadratic mixing also singles out a simpler and more natural basis for the cubic interactions.  In terms of this covariant derivative defined in (\ref{cov-D-t}), Eq.~\eqref{S3} can be rewritten as
\begin{align}
S_{\rm int}^{(3)} 
=  &  
\int_{x} \! a^3 
\bigg[ 
 \alpha  m_{\rm Pl} \sqrt{\frac{\epsilon}{2}} \,\sigma (D_t \zeta)^2 \nn 
\\
& \qquad  +  \frac{ \beta H }{2} \!
\sigma^2 D_t \zeta
+  \frac{\gamma H^2 } {3!  m_{\rm Pl} \sqrt{2 \epsilon} } \sigma^3
\bigg] ,
\label{S3-covariant}
\end{align}
where $\alpha=\widetilde{\alpha}$, $\beta=\widetilde{\beta}+2\lambda \widetilde{\alpha}$, and $\gamma=\widetilde{\gamma}+3\lambda \widetilde{\beta}+3\lambda^2\widetilde{\alpha}$. Thus, replacing $\dot\zeta$ with $D_t\zeta$ introduces no new cubic operators, but merely reshuffles their coefficients. 

The remainder of this article is organized as follows. We begin by deriving the exact linear solutions and fixing their normalization through the initial vacuum conditions. We then recover the exact power spectrum and use the mode functions to formulate dressed Schwinger--Keldysh rules, which we subsequently apply to the three interactions in Eq.~\eqref{S3-covariant}. Finally, we derive the corresponding squeezed-limit bispectra and analyze their behavior in the weak- and strong-mixing regimes.


\section{Linear dynamics}
\label{sec:linear-dynamics}

To study the dynamics of the system, it is convenient to introduce the canonically normalized curvature perturbation
\be
\varphi \equiv \sqrt{2\epsilon} \, m_{\rm Pl} \,  \zeta.
\label{canonical-z}
\ee
Neglecting slow-roll-suppressed corrections arising from the time dependence of $\epsilon$, the quadratic action in Eq.~\eqref{S2} becomes
\begin{align}
S^{(2)}
={}&
\frac12\int\dd[4]{x}\,a^3
\bigg[
\left(\dot\varphi-\lambda H\sigma\right)^2
-\frac{1}{a^2}(\nabla\varphi)^2
\nn\\
&
+\dot\sigma^2
-\frac{1}{a^2}(\nabla\sigma)^2
-\mu^2\sigma^2
\bigg].
\label{action-second-order}
\end{align}
It is useful to define the covariant derivative
\be
D_t\varphi
\equiv
\dot\varphi-\lambda H\sigma.
\label{cov-D_t}
\ee
This definition is related to the one introduced above by $D_t\varphi=\sqrt{2\epsilon} \, m_{\rm Pl} D_t\zeta$. Working to leading order in slow roll and taking $\lambda$ and $\mu/H$ to be constant, variation of Eq.~\eqref{action-second-order} yields the coupled linear equations of motion
\begin{align}
\frac{\dd}{\dd 
t}\left(D_t\varphi\right)
+3H D_t\varphi
-\frac{1}{a^2}\nabla^2\varphi
&=0,
\\
\ddot\sigma
+3H\dot\sigma
-\frac{1}{a^2}\nabla^2\sigma
+\mu^2\sigma
+\lambda H D_t\varphi
&=0.
\end{align}
The canonical momenta following from Eq.~\eqref{action-second-order} are $\Pi_\varphi=a^3D_t\varphi$ and $\Pi_\sigma=a^3\dot\sigma$. Together with the fields, they obey the equal-time commutation relations
\begin{align}
\Big[\varphi(\x,t),\Pi_\varphi(\x',t)\Big]
&=
i\delta^{(3)}(\x-\x'),
\\
\Big[\sigma(\x,t),\Pi_\sigma(\x',t)\Big]
&=
i\delta^{(3)}(\x-\x'),
\end{align}
with all other equal-time commutators vanishing. These relations are implemented by expanding the fields in Fourier space as
\begin{align}
\varphi(\x,t)
&=
\frac{1}{(2\pi)^3}
\int\dd[3]{k}\,
\varphi(\k,t)e^{i\k\cdot\x},
\label{quant-1}
\\
\sigma(\x,t)
&=
\frac{1}{(2\pi)^3}
\int\dd[3]{k}\,
\sigma(\k,t)e^{i\k\cdot\x},
\label{quant-2}
\end{align}
where the Fourier-space field operators are decomposed as
\begin{align}
\varphi(\k,t)
&=
\sum_b
\left[
\varphi_b(k,t)\hat a_b(\k)
+\varphi_b^*(k,t)\hat a_b^\dagger(-\k)
\right],
\label{Fourier-1}
\\
\sigma(\k,t)
&=
\sum_b
\left[
\sigma_b(k,t)\hat a_b(\k)
+\sigma_b^*(k,t)\hat a_b^\dagger(-\k)
\right].
\label{Fourier-2}
\end{align}
Here $b=1,2$ labels the two independent mode doublets, whose curvature and isocurvature components are $\varphi_b(k,t)$ and $\sigma_b(k,t)$, respectively. The creation and annihilation operators satisfy
\be
\Big[\hat a_b(\k),\hat a_c^\dagger(\k')\Big]
=
\delta_{bc}(2\pi)^3\delta^{(3)}(\k-\k').
\label{cre-ann}
\ee
Suppressing the label $b$, the equations governing the mode functions become
\begin{align}
\frac{\dd}{\dd t}\left(D_t\varphi\right)
+3H D_t\varphi
+\frac{k^2}{a^2}\varphi
&=0,
\\
\ddot\sigma
+3H\dot\sigma
+\frac{k^2}{a^2}\sigma
+\mu^2\sigma
+\lambda H D_t\varphi
&=0.
\end{align}
These equations can be simplified further by introducing the dimensionless variable
\be
z(t)\equiv\frac{k}{Ha(t)}.
\ee
Thus, $z$ is the physical wavenumber $k/a(t)$ measured in Hubble units. At early times, $z\gg1$ and the mode lies deep inside the Hubble radius; Hubble-radius crossing occurs at $z=1$; and at late times, $z\to0$ as the mode enters the superhorizon regime. At the order in slow roll considered here, $\dot z=-Hz$, and the time-domain covariant derivative can be written as $D_t\varphi=-HzD_z\varphi$. The mode equations then become
\begin{align}
\dv{z}\left(D_z\varphi\right)
-\frac{2}{z}D_z\varphi
+\varphi
&=0,
\label{eom-1}
\\
\dv[2]{\sigma}{z}
-\frac{2}{z}\dv{\sigma}{z}
+\left(
1+\frac{\mu^2}{H^2z^2}
\right)\sigma
&=
\frac{\lambda}{z}D_z\varphi,
\label{eom-2}
\end{align}
where
\be
D_z\varphi
\equiv
\dv{\varphi}{z}
+\frac{\lambda}{z}\sigma.
\label{cov-D_z}
\ee
Solving the coupled system in Eqs.~\eqref{eom-1} and~\eqref{eom-2} is the central analytic task of the following sections.


\section{Bogoliubov decomposition}
\label{sec:bogoliubov}

In the absence of quadratic mixing, $\lambda=0$, the canonical curvature perturbation $\varphi$ and the isocurvature perturbation $\sigma$ decouple. Choosing the two independent mode doublets introduced above, their mode functions are
\begin{align}
\varphi_b(k,z)
&=
\delta_{b1}u_0(k,z),
\\
\sigma_b(k,z)
&=
\delta_{b2}u_\mu(k,z),
\end{align}
where $u_0(k,z)$ and $u_\mu(k,z)$ are the Bunch--Davies mode functions for massless and massive scalars, respectively:
\begin{align}
u_0(k,z)
&=
\frac{iH}{\sqrt{2k^3}}
\left(1-iz\right)e^{iz},
\label{def-u-0}
\\
u_\mu(k,z)
&=
-\frac{H}{\sqrt{2k^3}}
\sqrt{\frac{\pi}{2}}\,
e^{i\theta_\nu}
z^{3/2}H_\nu^{(1)}(z).
\label{def-u-mu}
\end{align}
Here $H_\nu^{(1)}(z)$ is the Hankel function of the first kind, with index
\be
\nu=\sqrt{\frac94-\frac{\mu^2}{H^2}},
\ee
and $\theta_\nu=\frac{\pi}{2}\left(\nu-\frac32\right)$. The quantity $\theta_\nu$ fixes the Bunch--Davies normalization, such that for real $\nu$ it is a phase, while its analytic continuation provides the corresponding normalization for imaginary $\nu$. With this convention, both $u_0$ and $u_\mu$ approach the same positive-frequency mode $Hz\,e^{iz}/\sqrt{2k^3}$ as $z\to\infty$. Using these decoupled modes as a reference basis, we parametrize the fully coupled solutions as~\cite{Parra:2024usv}
\begin{align}
\varphi_b(k,z)
&=
f_b(z)u_0(k,z)
+\bar f_b(z)u_0^*(k,z),
\label{zeta-bogos}
\\
\sigma_b(k,z)
&=
s_b(z)u_\mu(k,z)
+\bar s_b(z)u_\mu^*(k,z).
\label{sigma-bogos}
\end{align}
The functions $f_b$, $\bar f_b$, $s_b$, and $\bar s_b$ are time-dependent Bogoliubov coefficients. The overbar labels the Bogoliubov partner and does not denote complex conjugation, which is indicated explicitly by an asterisk.

\subsection{Evolution of the Bogoliubov coefficients}

The coupled equations of motion \eqref{eom-1} and \eqref{eom-2} can be recast as a first-order system for the Bogoliubov coefficients. Ordering them as $(f_b,\bar f_b,s_b,\bar s_b)^T$ and suppressing the mode label $b$, one obtains~\cite{Fumagalli:2021mpc,Parra:2024usv}
\be
z\dv{z}
\left(
\begin{array}{c}
f\\
\bar f\\
s\\
\bar s
\end{array}
\right)
=
\left(
\begin{array}{cc}
0&\mathcal P(z)\\
\mathcal Q(z)&0
\end{array}
\right)
\left(
\begin{array}{c}
f\\
\bar f\\
s\\
\bar s
\end{array}
\right),
\label{first-order-eq-bogos}
\ee
where $\mathcal P(z)$ and $\mathcal Q(z)$ are $2\times2$ matrices given by
\begin{align}
\mathcal P(z)
&=
-i\frac{\lambda k^3}{H^2z^2}
\left(
\begin{array}{cc}
{u_0^*}'u_\mu&{u_0^*}'u_\mu^*\\
-u_0'u_\mu&-u_0'u_\mu^*
\end{array}
\right),
\\
\mathcal Q(z)
&=
-i\frac{\lambda k^3}{H^2z^2}
\left(
\begin{array}{cc}
u_0'u_\mu^*&{u_0^*}'u_\mu^*\\
-u_0'u_\mu&-{u_0^*}'u_\mu
\end{array}
\right).
\end{align}
In the previous expressions primes denote derivatives with respect to $z$. The factor $k^3/H^2$ cancels the corresponding normalization of the mode-function products, rendering both matrices independent of $k$. Thanks to these equations, the time dependent Bogoliubov coefficients satisfy the following two sets of constraints~\cite{Parra:2024usv}:
\begin{align}
\sum_b \Big( | f_b |^2 - | \bar f_b |^2 \Big) & = 1 , \label{bogo-constr-1} \\ 
\sum_b \Big( | s_b |^2 - | \bar s_b |^2 \Big) & = 1 , \\
\sum_b \Big( f_b s_b^*  - \bar f_b^* \bar s_b  \Big) & = 0 , \label{bogo-constr-3} 
\end{align}
and
\begin{align}
\sum_b \Big( f_b \bar s_b^*  - \bar f_b^*  s_b  \Big) & = 0 ,  \label{bogo-constr-4} \\
\sum_b \Big( f_b \bar f_b^*  - \bar f_b^*  f_b  \Big) & = 0 , \\
\sum_b \Big( s_b \bar s_b^*  - \bar s_b^*  s_b  \Big) & = 0  . \label{bogo-constr-6}
\end{align}
To select the Bunch--Davies vacuum, we may therefore impose initial conditions at a fiducial time $z_0\gg 1$ sufficiently deep in the ultraviolet. We choose
\begin{align}
(f_1,\bar f_1,s_1,\bar s_1)
&=
(1,0,0,0),
\label{init-1}
\\
(f_2,\bar f_2,s_2,\bar s_2)
&=
(0,0,1,0).
\label{init-2}
\end{align}
For $\lambda=0$, these initial conditions reproduce the decoupled Bunch--Davies solutions for $\varphi$ and $\sigma$.\\

\subsection{Fourth-order formulation}

We now reduce the coupled system to a single higher-order equation. Starting from Eq.~\eqref{first-order-eq-bogos} and assuming $\lambda\neq0$, three of the four Bogoliubov coefficients can be eliminated to obtain a fourth-order differential equation for any chosen coefficient. For definiteness, $f(z)$ satisfies
\begin{align}
\Bigg[
&
z^2\dv[4]{z}
+4z(1+iz)\dv[3]{z}
+\left(
\frac94-\nu^2+10iz-4z^2
\right)\dv[2]{z}
\nn\\
&
+\left(
\frac{5i}{2}-2i\nu^2-4z
\right)\dv{z}
-\lambda^2
\Bigg]f(z)
=0.
\label{4th-ode-general-nu}
\end{align}
The derivation of Eq.~\eqref{4th-ode-general-nu} relies on recurrence relations obeyed by the Hankel functions entering the reference modes. The first-order system also allows the remaining coefficients to be reconstructed directly from $f$ as:
\begin{align}
\bar f
&=
\widehat{\mathcal O}_{\bar f f} \, f ,
\label{zeta-zeta+}
\\
s
&=
\widehat{\mathcal O}_{s f} \, f,
\label{sigma+zeta+}
\\
\bar s
&=
\widehat{\mathcal O}_{\bar s f} \, f  ,
\label{sigma-zeta+}
\end{align}
where the corresponding differential operators are
\begin{widetext}
\begin{align}
\widehat{\mathcal O}_{\bar f f}
&\equiv
e^{2iz}
\left[
1
-\frac{2i}{\lambda^2}
\left(
z^2\dv[3]{z}
+2z(1+iz)\dv[2]{z}
+\left(
\frac14-\nu^2+2iz
\right)\dv{z}
\right)
\right],
\label{zeta-op}
\\
\widehat{\mathcal O}_{s f}
&\equiv
-\frac{ie^{-i\theta_\nu}}{\lambda}
\sqrt{\frac{\pi}{2}}\,
e^{iz}z^{1/2}
\Bigg[
zH_\nu^{(2)}(z)\dv[2]{z}
+
\left(
\left(\nu+\frac12+iz\right)H_\nu^{(2)}(z)
-zH_{\nu-1}^{(2)}(z)
\right)\dv{z}
\Bigg],
\label{sigma+op} \\
\widehat{\mathcal O}_{\bar s f}
&\equiv
\frac{ie^{i\theta_\nu}}{\lambda}
\sqrt{\frac{\pi}{2}}\,
e^{iz}z^{1/2}
\Bigg[
zH_\nu^{(1)}(z)\dv[2]{z}
+
\left(
\left(\nu+\frac12+iz\right)H_\nu^{(1)}(z)
-zH_{\nu-1}^{(1)}(z)
\right)\dv{z}
\Bigg].
\label{sigma-op}
\end{align}
\end{widetext}
Thus, solving Eq.~\eqref{4th-ode-general-nu} for $f$ determines all four Bogoliubov coefficients through Eqs.~\eqref{zeta-zeta+}--\eqref{sigma-zeta+}.


\section{General solution}

For $\lambda\neq0$, Eq.~\eqref{4th-ode-general-nu} provides an equivalent fourth-order formulation of the original first-order system~\eqref{first-order-eq-bogos}. On the interval $z\in(0,\infty)$, its solution space is four-dimensional, and the solutions derived below can subsequently be analytically continued to the complex $z$-plane on the appropriate branches. It is useful to introduce the second-order differential operator
\be
\widehat{\mathcal B}
\equiv
z\frac{\dd^2}{\dd z^2}
+
\left(1+2iz\right)\frac{\dd}{\dd z}.
\ee
The fourth-order equation~\eqref{4th-ode-general-nu} can then be written compactly as
\be
\widehat{\mathcal A}_\nu\,f=0,
\label{eq:Anu-zeta}
\ee
where
\be
\widehat{\mathcal A}_\nu
\equiv
\left(\widehat{\mathcal B}+\lambda\right)
\left(\widehat{\mathcal B}-\lambda\right)
+
\frac{1}{z}
\left(\frac14-\nu^2\right)
\left(
\widehat{\mathcal B}-\dv{z}
\right).
\label{def-Anu}
\ee
Before deriving the solution for arbitrary values of $\lambda$ and $\nu$, it is instructive to consider the conformal case $\nu=1/2$.

\subsection{Conformally coupled fields: $\nu=1/2$}

For $\nu=1/2$, the second term in Eq.~\eqref{def-Anu} vanishes and the fourth-order operator factorizes as
\be
\left(\widehat{\mathcal B}+\lambda\right)
\left(\widehat{\mathcal B}-\lambda\right)f=0.
\ee
For $\lambda\neq0$, the four-dimensional solution space therefore decomposes into two second-order sectors, labeled by $w=\pm1$, satisfying
\be
\left(\widehat{\mathcal B}-w\lambda\right)
f_{1/2}^{(w)}(z)=0,
\qquad
w=\pm1.
\label{eq:conformal-sector}
\ee
Thus, $w$ labels the two eigenvalue sectors $\pm\lambda$ of $\widehat{\mathcal B}$. Introducing the variable $x=-2iz$, Eq.~\eqref{eq:conformal-sector} becomes
\be
\left[
x\dv[2]{x}
+
(1-x)\dv{x}
-
\frac{iw\lambda}{2}
\right]f_{1/2}^{(w)}
=0,
\ee
which is Kummer's equation. For each value of $w$, two independent solutions are
\be
U\left(
\frac{iw\lambda}{2},1,-2iz
\right),
\qquad
M\left(
\frac{iw\lambda}{2},1,-2iz
\right),
\label{tricomi-kummer}
\ee
where $U(a,b,z)$ and $M(a,b,z)$ denote the Tricomi and Kummer confluent hypergeometric functions, respectively. Their relevant properties and analytic continuations are reviewed in Appendix~\ref{app:Tri-Kum}.

\subsection{Solutions for half-integer values of $\nu$}

The conformal solutions obtained above can be used as seeds to generate solutions for other half-integer values of $\nu$. To see this, we introduce the raising and lowering operators
\begin{align}
\widehat{\mathfrak R}_{\nu+1}
&\equiv
\left(\frac12+\nu\right)
+
i\left[
\widehat{\mathcal B}
-
\left(\frac12+\nu\right)\dv{z}
\right],
\label{def-raising}
\\
\widehat{\mathfrak L}_{\nu-1}
&\equiv
\left(\frac12-\nu\right)
+
i\left[
\widehat{\mathcal B}
-
\left(\frac12-\nu\right)\dv{z}
\right].
\label{def-lowering}
\end{align}
These operators satisfy the intertwining relations
\begin{align}
\widehat{\mathfrak R}_{\nu+1}\widehat{\mathcal A}_\nu
&=
\widehat{\mathcal A}_{\nu+1}
\widehat{\mathfrak R}_{\nu+1},
\label{eq:raising-intertwiner}
\\
\widehat{\mathfrak L}_{\nu-1}\widehat{\mathcal A}_\nu
&=
\widehat{\mathcal A}_{\nu-1}
\widehat{\mathfrak L}_{\nu-1}.
\label{eq:lowering-intertwiner}
\end{align}
On the graded space of $\nu$-sectors, they also obey the indexed commutation relation
\be
\widehat{\mathfrak R}_{\nu}
\widehat{\mathfrak L}_{\nu-1}
-
\widehat{\mathfrak L}_{\nu}
\widehat{\mathfrak R}_{\nu+1}
=
2\nu.
\ee
Together with the inclusion of a diagonal action $\widehat{\mathfrak N}f_\nu=\nu f_\nu$, this relation identifies $\widehat{\mathfrak R}$ and $\widehat{\mathfrak L}$ as the raising and lowering generators of an $\mathfrak{sl}(2)$ ladder algebra. Its quadratic Casimir acts on the $\nu$-sector as $-(\widehat{\mathcal A}_\nu+\lambda^2+\tfrac14)$. Consequently, if $f_\nu(z)$ satisfies $\widehat{\mathcal A}_\nu f_\nu=0$, then $\widehat{\mathfrak R}_{\nu+1}f_\nu$ and $\widehat{\mathfrak L}_{\nu-1}f_\nu$ solve the corresponding equations with indices $\nu+1$ and $\nu-1$, respectively:
\begin{align}
\widehat{\mathcal A}_{\nu+1}
\left(
\widehat{\mathfrak R}_{\nu+1}f_\nu
\right)
&=
0,
\\
\widehat{\mathcal A}_{\nu-1}
\left(
\widehat{\mathfrak L}_{\nu-1}f_\nu
\right)
&=
0.
\end{align}
Starting from either of the two conformal seeds in Eq.~\eqref{tricomi-kummer}, collectively denoted by $f_{1/2}^{(w)}(z)$, repeated applications of the raising operator generate a tower of solutions with half-integer $\nu$. These solutions may be written as
\be
f_{n+1/2}^{(w)}(z)
\equiv
\frac{1}{(1+iw\lambda)_n}
\widehat{\mathfrak R}_{n+1/2}
\cdots
\widehat{\mathfrak R}_{5/2}
\widehat{\mathfrak R}_{3/2}
f_{1/2}^{(w)}(z),
\label{f-from-g-1}
\ee
where $(\alpha)_n$ is the Pochhammer symbol, defined in terms of $\Gamma$-functions as:
\be
(\alpha)_n
\equiv
\frac{\Gamma(\alpha+n)}{\Gamma(\alpha)}.
\ee
For each of the two values of $w$ and each of the two conformal seeds, Eq.~\eqref{f-from-g-1} produces one of the four solutions to Eq.~\eqref{4th-ode-general-nu} at $\nu=n+1/2$, with $n=0,1,2,\ldots$. The ordered product is understood to be the identity when $n=0$. Acting on the conformal seeds with the lowering operator produces no additional linearly independent solutions, as expected.

\subsection{Solutions for general values of $\nu$}

The ladder construction suggests that solutions for arbitrary $\nu$ may be generated by a $\nu$-dependent function of a differential operator acting on the conformal seeds. We therefore introduce
\be
\hat x
\equiv
\frac{i}{2}\dv{z},
\label{def-p-hat}
\ee
and consider the ansatz
\be
f_\nu^{(w)}(z)
=
D_\nu^{(w)}(\hat x)\,
f_{1/2}^{(w)}(z),
\label{eq:general-ansatz}
\ee
where $D_\nu^{(w)}(x)$ is an ordinary function of its argument and $f_{1/2}^{(w)}$ denotes either conformal seed. Equivalently, in a representation that diagonalizes $\hat x$, the ansatz takes the multiplicative form $\widetilde f_\nu^{(w)}(x)=D_\nu^{(w)}(x)\widetilde f_{1/2}^{(w)}(x)$. We impose the normalization $ D_\nu^{(w)}(0)=1$, which agrees with that of the ladder-generated solutions in Eq.~\eqref{f-from-g-1}. To determine $D_\nu^{(w)}(x)$, we first note that $\widehat{\mathcal B}$ and $\hat x$ satisfy
\be
\left[
\widehat{\mathcal B},\hat x
\right]
=
-2i\,\hat x(1-\hat x).
\label{eq:B-p-commutator}
\ee
Using Eq.~\eqref{eq:conformal-sector}, one then finds that the action of $\widehat{\mathcal A}_\nu$ on the ansatz~\eqref{eq:general-ansatz} is
\begin{widetext}
\be
\widehat{\mathcal A}_\nu
\Big(
D_\nu^{(w)}(\hat x)f_{1/2}^{(w)}
\Big)
=
-4\hat x(1-\hat x)
\bigg[
\hat x(1-\hat x)
D_\nu^{(w)}{}''(\hat x)
+
\left(
1+iw\lambda-2\hat x
\right)
D_\nu^{(w)}{}'(\hat x)
-
\left(
\frac14-\nu^2
\right)
D_\nu^{(w)}(\hat x)
\bigg]
f_{1/2}^{(w)}.
\label{eq:Anu-on-Fg}
\ee
\end{widetext}
Here, primes denote derivatives of the ordinary function $D_\nu^{(w)}(x)$ with respect to its argument, followed by the replacement $x\to\hat x$. These manipulations follow most transparently in a representation in which $\hat x$ acts multiplicatively. It follows that the ansatz~\eqref{eq:general-ansatz} solves the fourth-order equation provided that $D_\nu^{(w)}(x)$ satisfies
\be
x(1-x)D''
+
\left(
1+iw\lambda-2x
\right)D'
-
\left(
\frac14-\nu^2
\right)D
=
0.
\label{eq:F-hypergeometric}
\ee
This is precisely the Gauss hypergeometric equation with parameters $a=\frac12-\nu$, $b=\frac12+\nu$, and $c=1+iw\lambda$. The solution analytic at $x=0$, and normalized according to $ D_\nu^{(w)}(0)=1$, is therefore
\be
D_\nu^{(w)}(x)
=
{}_2F_1\left(
\frac12-\nu,\frac12+\nu;
1+iw\lambda;
x
\right),
\label{eq:F-solution}
\ee
where ${}_2F_1(a,b;c;x)$ denotes the Gauss hypergeometric function. This result motivates the operator definition
\be
\widehat{\mathcal D}_\nu^{(w)}
\equiv
{}_2F_1\left(
\frac12-\nu,\frac12+\nu;
1+iw\lambda;
\frac{i}{2}\dv{z}
\right),
\label{def-Dnu-w}
\ee
in terms of which the ansatz becomes
\be
f_\nu^{(w)}(z)
\equiv
\widehat{\mathcal D}_\nu^{(w)}
f_{1/2}^{(w)}(z).
\label{def-fnujs}
\ee
Formally, the operator in Eq.~\eqref{def-Dnu-w} is represented by the hypergeometric series, with its final argument acting as a differential operator:
\be
\widehat{\mathcal D}_\nu^{(w)}
=
\sum_{k=0}^{\infty}
\frac{
\left(\frac12-\nu\right)_k
\left(\frac12+\nu\right)_k
}{
(1+iw\lambda)_k\,k!
}
\left(
\frac{i}{2}\dv{z}
\right)^k.
\label{eq:Dnu-series}
\ee
For $\nu=1/2$, one of the upper hypergeometric parameters vanishes, so $\widehat{\mathcal D}_{1/2}^{(w)}$ reduces to the identity and the conformal solutions~\eqref{tricomi-kummer} are recovered directly.

For half-integer values $\nu=n+1/2$, with $n=0,1,2,\ldots$, the first upper parameter in Eq.~\eqref{eq:Dnu-series} is $-n$. The series therefore terminates, and $\widehat{\mathcal D}_{n+1/2}^{(w)}$ becomes a differential operator of order $n$:
\be
\widehat{\mathcal D}_{n+1/2}^{(w)}
=
\sum_{k=0}^{n}
\frac{
(-n)_k(n+1)_k
}{
(1+iw\lambda)_k\,k!
}
\left(
\frac{i}{2}\dv{z}
\right)^k.
\label{eq:finite-Dnu}
\ee
Consistency with the ladder construction follows from the contiguous relations of the Gauss hypergeometric function. In particular, the solutions~\eqref{def-fnujs} satisfy
\begin{align}
\widehat{\mathfrak R}_{\nu+1}
f_\nu^{(w)}
&=
\left(
\frac12+\nu+iw\lambda
\right)
f_{\nu+1}^{(w)},
\label{eq:raising-action}
\\
\widehat{\mathfrak L}_{\nu-1}
f_\nu^{(w)}
&=
\left(
\frac12-\nu+iw\lambda
\right)
f_{\nu-1}^{(w)}.
\label{eq:lowering-action}
\end{align}
Iterating Eq.~\eqref{eq:raising-action} from $\nu=1/2$ gives
\be
\widehat{\mathcal D}_{n+1/2}^{(w)}
f_{1/2}^{(w)}
=
\frac{1}{(1+iw\lambda)_n}
\widehat{\mathfrak R}_{n+1/2}
\cdots
\widehat{\mathfrak R}_{5/2}
\widehat{\mathfrak R}_{3/2}
f_{1/2}^{(w)},
\label{eq:Dnu-ladder-product}
\ee
in agreement with Eq.~\eqref{f-from-g-1}. Equivalently, Eq.~\eqref{eq:Dnu-ladder-product} can be established directly by induction on $n$.

\subsection{The four independent solutions}

Collecting the preceding results, it is convenient to write the four independent solutions to Eq.~\eqref{4th-ode-general-nu} as
\begin{align}
f_U^{(w)}(z)
&=
\frac{e^{\pi w\lambda/4}}{\sqrt{2}}\,
\widehat{\mathcal D}_\nu^{(w)}
U\left(
\frac{iw\lambda}{2},1,-2iz
\right),
\label{f-sol-1}
\\
f_M^{(w)}(z)
&=
N_w  \bigg[ \frac{e^{-\pi w\lambda/4}}{\sqrt{2}}\,
\widehat{\mathcal D}_\nu^{(w)}
M\left(
\frac{iw\lambda}{2},1,-2iz
\right) \nn \\ 
&\quad
-   \frac{1}{ \Gamma(1 - \frac{i w \lambda}{2})} f_U^{(w)}(z) \bigg],
\label{f-sol-2}
\end{align}
where the normalization coefficient $N_w$ is defined as:
\be
N_w  \equiv  \frac{\Gamma (\frac{i w \lambda}{2})}{2 \Gamma^2 (i w \lambda)} \Gamma \Big(\frac{1}{2} + \nu + i w \lambda \Big) \Gamma \Big(\frac{1}{2} - \nu + i w \lambda \Big) .
\ee
For generic $\lambda\neq0$, the four functions in Eqs.~\eqref{f-sol-1} and~\eqref{f-sol-2} form a linearly independent basis. The general solution associated with each independent mode $b$ can therefore be written as
\be
f_b(z)
=
\sum_{w=\pm1}
\left[
A_b^{(w)}f_U^{(w)}(z)
+
B_b^{(w)}f_M^{(w)}(z)
\right],
\label{general-sol-A}
\ee
where the coefficients $A_b^{(w)}$ and $B_b^{(w)}$ are fixed by the initial conditions~\eqref{init-1} and~\eqref{init-2}. The remaining Bogoliubov coefficients, $\bar f_b(z)$, $s_b(z)$, and $\bar s_b(z)$, are reconstructed from $f_b(z)$ using Eqs.~\eqref{zeta-zeta+}--\eqref{sigma-zeta+}. One obtains
\begin{align}
\bar f_b(z)
&=  
\sum_{w=\pm1} 
\Big[
A_b^{(w)}
\bar f_U^{(w)}(z)
+
B_b^{(w)} \bar f_M^{(w)}(z)
\Big],
\label{zeta-f}
\\
s_b(z)
&= 
\sum_{w=\pm1}
\Big[
A_b^{(w)} s_U^{(w)}(z) 
+ 
B_b^{(w)} s_M^{(w)}(z) 
\Big],
\label{sigma+f}
\\
\bar s_b(z)
&=
\sum_{w=\pm1} 
\Big[
A_b^{(w)} \bar s_U^{(w)}(z) 
+
B_b^{(w)} \bar s_M^{(w)}(z)
\Big],
\label{sigma-f}
\end{align}
where $\bar f_X^{(w)}(z)$, $s_X^{(w)}(z)$ and $\bar s_X^{(w)}(z)$, with $X = \{ U, M \}$, are naturally defined as
\begin{align}
\bar f_X^{(w)}(z) & = \widehat{\mathcal O}_{\bar f f}
 f_X^{(w)}(z) ,  \\
s_X^{(w)}(z) & = \widehat{\mathcal O}_{s f}
 f_X^{(w)}(z) ,  \\
 \bar s_X^{(w)}(z) & = \widehat{\mathcal O}_{\bar s f}
 f_X^{(w)}(z) ,
\end{align}
where $\widehat{\mathcal O}_{\bar f f}$, $\widehat{\mathcal O}_{s f}$, and $\widehat{\mathcal O}_{\bar s f}$ are the differential operators introduced in Eqs.~\eqref{zeta-op}, \eqref{sigma+op}, and~\eqref{sigma-op}, respectively. 

To finish, notice that with the normalization chosen in Eqs.~\eqref{f-sol-1} and~\eqref{f-sol-2}, the algebraic constraints~\eqref{bogo-constr-1}--\eqref{bogo-constr-3} leads to the following relation satisfied by the integration constants introduced in (\ref{general-sol-A}):
\be
\sum_{b=1}^{2}
\left(
A_b^{(w)}A_b^{(w')*}
-
B_b^{(w)}B_b^{(w')*}
\right)
=
\delta_{ww'}.
\label{AA-BB-cond}
\ee
On the other hand, thanks to the second set of constraints ~\eqref{bogo-constr-4}--\eqref{bogo-constr-6}, one additionally finds
\be
\sum_{b=1}^{2}
\left(
A_b^{(w)}B_b^{(w')*}
-
B_b^{(w)*}A_b^{(w')}
\right)
=
0.
\label{AB-BA-cond}
\ee
In Appendix~\ref{app:inner-product} we offer an alternative derivation of these relations based on the mode functions to be introduced in the next section.


\section{Mode-function representations}

To formulate the Schwinger--Keldysh diagrammatic rules used to compute $n$-point correlation functions, which will be developed in Section~\ref{sec:SK-rules}, it is useful to obtain explicit representations of the mode functions $\varphi_b$ and $\sigma_b$ in terms of the Tricomi and Kummer functions.

\subsection{Full mode reconstruction}

Using Eqs.~\eqref{zeta-f}--\eqref{sigma-f}, the physical mode functions~\eqref{zeta-bogos} and~\eqref{sigma-bogos} can be written as
\begin{align}
\varphi_b(k,z)
&= \!\!\!
\sum_{w=\pm1} \!\!
\left[ 
A_b^{(w)}\mathcal F_U^{(w)}(k,z)
+
B_b^{(w)}\mathcal F_M^{(w)}(k,z)
\right],
\label{varphi-F}
\\
\sigma_b(k,z)
&= \!\!\!
\sum_{w=\pm1} \!\!
\left[ 
A_b^{(w)}\mathcal S_U^{(w)}(k,z)
+
B_b^{(w)}\mathcal S_M^{(w)}(k,z)
\right],
\label{s-F}
\end{align}
where, for $X\in\{U,M\}$, we have defined
\begin{align}
\mathcal F_X^{(w)}(k,z)
&\equiv
u_0(k,z) f_X^{(w)}(z) + 
u_0^*(k,z) \bar
 f_X^{(w)}(z) ,
\label{def-ZsX}
\\
\mathcal S_X^{(w)}(k,z)
&\equiv u_\mu(k,z)
 s_X^{(w)}(z)  
+
u_\mu^*(k,z) \bar
 s_X^{(w)}(z) .
\label{def-SsX}
\end{align}
Owing to the properties of the Tricomi and Kummer functions, these particular combinations of the Bunch--Davies mode functions $u_0$ and $u_\mu$ with the solutions~\eqref{f-sol-1} and~\eqref{f-sol-2} admit a considerable simplification.

To exhibit this simplification, we introduce the auxiliary functions
\begin{align}
g_U^{(w)}(z)
&\equiv
-\frac{e^{\frac{\pi w\lambda}{4}}}{\sqrt{2}}\,
\widehat{\mathcal D}_\nu^{(w)} \!\!
\left[
2iz\,
U\left(
1+\frac{iw\lambda}{2},
2,-2iz
\right)
\right],
\label{eq-CUs-hypergeometric}
\\
g_M^{(w)}(z)
&\equiv
N_w \Bigg[
\frac{e^{-\frac{\pi w\lambda}{4}}}{\sqrt{2}}\,
\widehat{\mathcal D}_\nu^{(w)} \!\!
\left[
2iz\,
M\left(
1+\frac{iw\lambda}{2},
2,-2iz
\right)
\right] \nn \\
& \quad
-  \frac{1}{ \Gamma(1 - \frac{i w \lambda}{2})} g_U^{(w)}(z) \Bigg] ,
\label{eq-CMs-hypergeometric}
\end{align}
\begin{align}
h_U^{(w)}(z)
&\equiv
\frac{e^{\frac{\pi w\lambda}{4}}}{\sqrt{2}}\,
\widehat{\mathcal D}_\nu^{(w)}
U\left(
1+\frac{iw\lambda}{2},
2,-2iz
\right),
\label{def-TUs}
\\
h_M^{(w)}(z)
&\equiv
-  N_w \Bigg[ \frac{e^{-\frac{\pi w\lambda}{4}}}{\sqrt{2}}\,
\widehat{\mathcal D}_\nu^{(w)}
M\left(
1+\frac{iw\lambda}{2},
2,-2iz
\right) \nn \\
& \quad
+  \frac{1}{ \Gamma(1 - \frac{i w \lambda}{2})} h_U^{(w)}(z) \Bigg] .
\label{def-TMs}
\end{align}
These functions satisfy the simple relations
\be
\frac{i}{2}\dv{z}g_X^{(w)}
=
g_X^{(w)} -f_X^{(w)},
\label{eq-Cf-relation}
\ee
and
\be
\dv{z}f_X^{(w)}
=
-w\lambda h_X^{(w)}.
\label{eq-fT-relation}
\ee
In terms of these functions, and thanks to the properties listed in Appendix~\ref{app:Tri-Kum}, the combinations defined in Eqs.~\eqref{def-ZsX} and~\eqref{def-SsX} reduce to
\begin{align}
\mathcal F_X^{(w)}(k,z)
&=
\frac{He^{iz}}{\sqrt{2k^3}}
\left[
2zf_X^{(w)}(z)
+
i(1+iz)g_X^{(w)}(z)
\right],
\label{eq-Z-fC}
\\
\mathcal S_X^{(w)}(k,z)
&=
\frac{H}{\sqrt{2k^3}}\,
2wz^2e^{iz}h_X^{(w)}(z).
\label{eq-S-T}
\end{align}
Applying the covariant derivative $D_z$ to Eq.~\eqref{eq-Z-fC} and using Eqs.~\eqref{eq-Cf-relation} and~\eqref{eq-fT-relation} gives an equally simple expression for the covariant derivative of the curvature mode:
\begin{align}
D_z\mathcal F_X^{(w)}(k,z)
&\equiv
\dv{z}\mathcal F_X^{(w)}(k,z)
+
\frac{\lambda}{z}\mathcal S_X^{(w)}(k,z)
\nn\\
&=
\frac{H}{\sqrt{2k^3}}\,
ize^{iz}g_X^{(w)}(z).
\label{eq-Dz-Z}
\end{align}
These representations will considerably simplify the Schwinger--Keldysh rules used below to compute $n$-point correlation functions.

\subsection{Integral representations}

The definitions~\eqref{f-sol-1}--\eqref{f-sol-2} and~\eqref{eq-CUs-hypergeometric}--\eqref{def-TMs} involve the action of $\widehat{\mathcal D}_\nu^{(w)}$ on Tricomi and Kummer functions. We now derive integral representations that make this action explicit. For brevity, we focus on the modes generated by the Tricomi function $U$; the construction for the Kummer-generated modes proceeds analogously.

We begin by defining the conformal seed functions
\begin{align}
f_{Uc}^{(w)}(z)
&\equiv
\frac{e^{\pi w\lambda/4}}{\sqrt{2}}
U\left(
\frac{iw\lambda}{2},1,-2iz
\right),
\label{f-seed}
\\
g_{Uc}^{(w)}(z)
&\equiv
\frac{e^{\pi w\lambda/4}}{\sqrt{2}}\,
U\left(
\frac{iw\lambda}{2},0,-2iz
\right),
\label{g-seed}
\\
h_{Uc}^{(w)}(z)
&\equiv
\frac{e^{\pi w\lambda/4}}{\sqrt{2}}
U\left(
1+\frac{iw\lambda}{2},2,-2iz
\right).
\label{h-seed}
\end{align}
The corresponding solutions at general $\nu$ are
\begin{align}
f_U^{(w)}(z)
&=
\widehat{\mathcal D}_\nu^{(w)}
f_{Uc}^{(w)}(z),
\\
g_U^{(w)}(z)
&=
\widehat{\mathcal D}_\nu^{(w)}
g_{Uc}^{(w)}(z),
\label{gu-guc}
\\
h_U^{(w)}(z)
&=
\widehat{\mathcal D}_\nu^{(w)}
h_{Uc}^{(w)}(z).
\end{align}
Equation~\eqref{gu-guc} agrees with Eq.~\eqref{eq-CUs-hypergeometric} by virtue of the identity~\eqref{U_id_b=2}. Using the integral representation~\eqref{app-int-U}, the conformal seeds~\eqref{f-seed}--\eqref{h-seed} become
\begin{align}
f_{Uc}^{(w)}(z)
&=
\frac{e^{\pi w\lambda/4}}
{\sqrt{2}\,\Gamma(\frac{iw\lambda}{2})}
\int_0^\infty
\frac{\dd{t}}{t}\,
e^{2izt} \!
\left(
\frac{t}{1+t}
\right)^{\frac{iw\lambda}{2}} \!\! ,
\label{f-seed-2}
\\
g_{Uc}^{(w)}(z)
&=
\frac{e^{\pi w\lambda/4}}
{\sqrt{2}\,\Gamma(\frac{iw\lambda}{2})}
\int_0^\infty
\frac{\dd{t}}{t}\,
\frac{e^{2izt}}{1+t} \!
\left(
\frac{t}{1+t}
\right)^{\frac{iw\lambda}{2}} \!\! ,
\label{g-seed-2}
\\
h_{Uc}^{(w)}(z)
&=
\frac{e^{\pi w\lambda/4}}
{\sqrt{2}\,\Gamma(1 \! + \! \frac{iw\lambda}{2})}
\int_0^\infty \!\!\!
\dd{t}\,
e^{2izt} \!
\left(
\frac{t}{1+t}
\right)^{\frac{iw\lambda}{2}} \!\!\! .
\label{h-seed-2}
\end{align}
These representations are initially valid in their common convergence domain, which includes $\operatorname{Re}(iw\lambda)>0$ and $\operatorname{Im}z>0$. For real $\lambda$, they are understood by analytic continuation in the parameters, with the positive real $z$ axis approached using the corresponding $i\epsilon$ prescription.

Since $e^{2izt}$ is an eigenfunction of $\hat x=\frac{i}{2}\dv{z}$ with eigenvalue $-t$, the spectral action of $\widehat{\mathcal D}_\nu^{(w)}$ inside these integrals amounts to multiplication by $D_\nu^{(w)}(-t) = {}_2F_1\left(
\frac12-\nu,\frac12+\nu;
1+iw\lambda;
-t
\right)$. We therefore obtain
\begin{align}
f_{U}^{(w)}(z)
=&
\frac{e^{\pi w\lambda/4}}
{\sqrt{2}\,\Gamma(\frac{iw\lambda}{2})}
\int_0^\infty
\frac{\dd{t}}{t}\,
e^{2izt} \!
\left(
\frac{t}{1+t}
\right)^{\frac{iw\lambda}{2}} \!\! \nn \\
& \times {}_2F_1\left(
\frac12-\nu,\frac12+\nu;
1+iw\lambda;
-t
\right) ,
\label{f-int-2}
\\
g_{U}^{(w)}(z)
&=
\frac{e^{\pi w\lambda/4}}
{\sqrt{2}\,\Gamma(\frac{iw\lambda}{2})}
\int_0^\infty
\frac{\dd{t}}{t}\,
\frac{e^{2izt}}{1+t} \!
\left(
\frac{t}{1+t}
\right)^{\frac{iw\lambda}{2}} \!\!  \nn \\
& \times {}_2F_1\left(
\frac12-\nu,\frac12+\nu;
1+iw\lambda;
-t
\right) ,
\label{g-int-2}
\\
h_{U}^{(w)}(z)
&=
\frac{e^{\pi w\lambda/4}}
{\sqrt{2}\,\Gamma(1 \! + \! \frac{iw\lambda}{2})}
\int_0^\infty \!\!\!
\dd{t}\,
e^{2izt} \!
\left(
\frac{t}{1+t}
\right)^{\frac{iw\lambda}{2}} \!\!\!  \nn \\
& \times {}_2F_1\left(
\frac12-\nu,\frac12+\nu;
1+iw\lambda;
-t
\right) .
\label{h-int-2}
\end{align}
These integral representations provide the desired realization of the formal differential operator introduced in the preceding section.

\subsection{Small-$z$ expansions}
\label{sec:small-z}

The small-$z$ behavior of the functions~\eqref{f-int-2}--\eqref{h-int-2} can be extracted from their integral representations. In particular, their nonanalytic contributions are controlled by the large-$t$ tail of the integrals, with the relevant integration region scaling as $t\sim z^{-1}$. For generic $2\nu\notin\mathbb Z$, the hypergeometric function $D_\nu^{(w)}(-t)$ has the large-$t$ expansion
\begin{align}
D_\nu^{(w)}(-t)
&=
C_+^{(w)}
t^{\nu-\frac12}
\left[
1+\mathcal O\left(t^{-1}\right)
\right]
\nn\\
&\quad+
C_-^{(w)}
t^{-\nu-\frac12}
\left[
1+\mathcal O\left(t^{-1}\right)
\right],
\label{F-large-t}
\end{align}
where
\be
C_\pm^{(w)}
=
\frac{
\Gamma(1+iw\lambda)\Gamma(\pm2\nu)
}{
\Gamma(\frac12\pm\nu)
\Gamma(\frac12\pm\nu+iw\lambda)
}.
\label{Cpm-def}
\ee
The two scaling exponents in Eq.~\eqref{F-large-t} are determined solely by $\nu$, whereas their coefficients depend on both $\nu$ and $\lambda$. At values for which $2\nu\in\mathbb Z$, Eq.~\eqref{F-large-t} is understood by taking the appropriate limit; depending on the degeneracy, logarithmic terms may then appear.

Using this asymptotic behavior, one can extract the leading nonanalytic contributions to the more directly relevant functions $\mathcal F_U^{(w)}(k,z)$ and $\mathcal S_U^{(w)}(k,z)$ defined in Eqs.~\eqref{eq-Z-fC} and~\eqref{eq-S-T}. For generic $0<\operatorname{Re}\nu<3/2$, the curvature mode has the expansion
\begin{align}
\mathcal F_U^{(w)} \! (k,z) 
&= \!
\frac{H}{\sqrt{2k^3}} \!
\Big[
\mathcal F_0^{(w)} \!
+ \!
\mathcal F_+^{(w)}
z^{\frac32-\nu} \!
+ \!
\mathcal F_-^{(w)}
z^{\frac32+\nu}
+
\cdots
\Big],
\label{ZU-small-z-general}
\end{align}
where $\mathcal F_0^{(w)}=ig_U^{(w)}(0)$ is given explicitly by
\be
\mathcal F_0^{(w)}
=
\frac{ie^{\pi w\lambda/4}}{\sqrt{2\pi}}
\frac{
\Gamma(\frac12+\frac{iw\lambda}{2})
\Gamma(\frac34-\frac{\nu}{2})
\Gamma(\frac34+\frac{\nu}{2})
}{
\Gamma(\frac34-\frac{\nu}{2}+\frac{iw\lambda}{2})
\Gamma(\frac34+\frac{\nu}{2}+\frac{iw\lambda}{2})
}.
\label{g0-integral-short}
\ee
On the other hand, the coefficients of the two nonanalytic branches are
\be
\mathcal F_\pm^{(w)} \!
=
\frac{
\sqrt{2}\,e^{\pi w\lambda/4} \left(
1 \mp 2 \nu
\right) 
}{
2 \Gamma(\frac{iw\lambda}{2})
} C_\pm^{(w)}
(-2i)^{\frac12\mp\nu}
\Gamma\left(
\frac32 \mp \nu
\right) \! .
\label{Qpm-def-alt}
\ee
For the same range of $\nu$, the entropy mode behaves as
\begin{align}
\mathcal S_U^{(w)}(k,z)
&=
\frac{H}{\sqrt{2k^3}}
\Big[
\mathcal S_+^{(w)}
z^{\frac32-\nu}
+
\mathcal S_-^{(w)}
z^{\frac32+\nu}
+
\cdots
\Big],
\label{SU-small-z-general}
\end{align}
where
\be
\mathcal S_\pm^{(w)}
=
\frac{
\sqrt{2}\,w e^{\pi w\lambda/4}
}{
\Gamma(1+\frac{iw\lambda}{2})
}
C_\pm^{(w)}
\Gamma\left(
\frac12\pm\nu
\right)
(-2i)^{\mp\nu-\frac12}.
\label{Rpm-def}
\ee
In Eqs.~\eqref{ZU-small-z-general} and~\eqref{SU-small-z-general}, the ellipses include analytic contributions and subleading terms within each nonanalytic branch. The phases of the complex powers are fixed by the same branch prescription used in the integral representations.

The expressions~\eqref{ZU-small-z-general} and~\eqref{SU-small-z-general} can be analytically continued under $\nu\to i\rho$ when the entropy mass enters the heavy regime, $\mu>3H/2$. At the degenerate value $\nu=3/2$, the leading curvature mode instead acquires a logarithmic dependence on $z$, while the entropy mode approaches a constant:
\begin{align}
\mathcal F_U^{(w)}(k,z)
&=
\frac{H}{\sqrt{2k^3}}
\left[
-\frac{
\sqrt{2}\,ie^{\pi w\lambda/4}
\log(-2iz) }{
(1+iw\lambda)
\Gamma(\frac{iw\lambda}{2})
}
+
\cdots
\right],
\label{ZU-small-z-nu-three-half}
\\
\mathcal S_U^{(w)}(k,z)
&=
\frac{H}{\sqrt{2k^3}}
\left[
-\frac{
w e^{\pi w\lambda/4}
}{
\sqrt{2}\,(1+iw\lambda)
\Gamma(1+\frac{iw\lambda}{2})
}
+
\cdots
\right].
\label{SU-small-z-nu-three-half}
\end{align}
These expressions determine the mode functions required below on superhorizon scales.


\section{Initial conditions}

We now determine the integration constants $A_b^{(w)}$ and $B_b^{(w)}$ introduced in Eq.~\eqref{general-sol-A} by imposing the initial conditions~\eqref{init-1} and~\eqref{init-2}. We begin by defining:
\begin{align}
\chi_w(z)
&\equiv
\frac{1}{\sqrt{2}}
\exp\left[
-\frac{iw\lambda}{2}\ln(2z)
\right] ,
\end{align}
which is a factor that recur in large-$z$ expansions. Up to corrections of order $\mathcal O(z^{-2})$, the normalized conformal seeds entering $f_U^{(w)}$ and $f_M^{(w)}$ behave as
\begin{align}
\frac{e^{\frac{\pi w\lambda}{4}}}{\sqrt{2}}\,
U\left(
\frac{iw\lambda}{2},1,-2iz
\right)
&\sim
\left[
1+\frac{i\lambda^2}{8z}
\right]\chi_w(z),
\\
\frac{e^{-\frac{\pi w\lambda}{4}}}{\sqrt{2}}\,
M\left(
\frac{iw\lambda}{2},1,-2iz
\right)
&\sim
\left[
1+\frac{i\lambda^2}{8z}
\right]\frac{\chi_w(z)}{
\Gamma\left(1-\frac{iw\lambda}{2}\right)
} 
\nn\\
& \!\!\!\! +
\frac{ie^{-2iz}}{
2z\,\Gamma\left(\frac{iw\lambda}{2}\right)
}
\chi_{-w}(z).
\end{align}
The Kummer-generated seed therefore contains both nonoscillatory and oscillatory branches, whereas the Tricomi-generated seed contains only the former.

The asymptotic expansions of $f_U^{(w)}$ and $f_M^{(w)}$ follow by applying $\widehat{\mathcal D}_\nu^{(w)}$ to these individual branches. For the nonoscillatory terms, only the first-derivative contribution in the series~\eqref{eq:Dnu-series} is needed at order $z^{-1}$. Indeed,
\be
\frac{i}{2}\dv{z}\chi_w(z)
=
\frac{w\lambda}{4z}\chi_w(z),
\ee
and hence
\be
\widehat{\mathcal D}_\nu^{(w)}
\chi_w(z)
=
\chi_w(z)
\left[
1+
\frac{w\lambda}{4z}
\frac{\frac14-\nu^2}{1+iw\lambda}
\right]
+
\mathcal O(z^{-2}).
\ee
The same relation holds with $\chi_w(z)$ replaced by $\psi_w(z)$. For the oscillatory branch, by contrast,
\be
\frac{i}{2}\dv{z}
\left[
e^{-2iz}\chi_{-w}(z)
\right]
=
\left[
1+\mathcal O(z^{-1})
\right]
e^{-2iz}\chi_{-w}(z).
\ee
Since this branch already carries an overall factor of $z^{-1}$, the correction displayed above contributes only at order $z^{-2}$. To the required accuracy, $\widehat{\mathcal D}_\nu^{(w)}$ therefore acts on the oscillatory factor by evaluating its hypergeometric argument at unity:
\begin{align}
\widehat{\mathcal D}_\nu^{(w)}
\left[
e^{-2iz}\chi_{-w}(z)
\right]
&=
\Bigg[
{}_2F_1\left(
\frac12-\nu,\frac12+\nu;
1+iw\lambda;
1
\right)
\nn\\
&\qquad+
\mathcal O(z^{-1})
\Bigg]
e^{-2iz}\chi_{-w}(z).
\end{align}
The hypergeometric factor can be simplified using the Gauss summation formula with $a= \frac12-\nu$, $b = \frac12+\nu$ and $c= 1+iw\lambda$:
\be
 {}_2F_1\left(
a, b; c;
1
\right)  =
\frac{
\Gamma(1+iw\lambda)\Gamma(iw\lambda)
}{
\Gamma\left(\frac12+\nu+iw\lambda\right)
\Gamma\left(\frac12-\nu+iw\lambda\right)
}.
\ee
Here $c-a-b=iw\lambda$ has vanishing real part. The value at unity is therefore understood through the analytic continuation in parameter space of the Gauss formula from the region $\operatorname{Re}(c-a-b)>0$.

Combining these results with the large-$z$ expansions of the normalized Tricomi and Kummer seeds gives
\begin{align}
f_U^{(w)}(z)
&=
\chi_w(z)
\left[
1+
\frac{1}{z}
\left(
\frac{i\lambda^2}{8}
+
\frac{w\lambda}{4}
\frac{\frac14-\nu^2}{1+iw\lambda}
\right)
\right]
\! \nn\\ 
&\quad +
\mathcal O(z^{-2}),
\\
f_M^{(w)}(z)
&=
-\frac{w\lambda}{4z}\,
\chi_{-w}(z)
+
\mathcal O(z^{-2}) .
\end{align}
Using the reconstruction operators~\eqref{zeta-op}--\eqref{sigma-op}, the Bogoliubov coefficients consequently have the following leading asymptotic behavior:
\begin{align}
f_b(z)
&=
\sum_{w=\pm1}
A_b^{(w)}\chi_w(z)
+
\mathcal O(z^{-1}),
\label{eq-zeta-plus-asymptotic}
\\
\bar f_b(z)
&=
\sum_{w=\pm1}
B_b^{(w)}\chi_{-w}(z)
+
\mathcal O(z^{-1}),
\label{eq-zeta-minus-asymptotic}
\\
s_b(z)
&=
i\sum_{w=\pm1}
w A_b^{(w)}\chi_w(z)
+
\mathcal O(z^{-1}),
\label{eq-sigma-plus-asymptotic}
\\
\bar s_b(z)
&=
-i\sum_{w=\pm1}
w B_b^{(w)}\chi_{-w}(z)
+
\mathcal O(z^{-1}).
\label{eq-sigma-minus-asymptotic}
\end{align}

We now evaluate these expressions at $z=z_0$ and impose the initial conditions~\eqref{init-1} and~\eqref{init-2}. The equations for $\bar f_b$ and $\bar s_b$ imply that $B_1^{(w)}$ and $B_2^{(w)}$ are at most of order $\mathcal O(z_0^{-1})$. Solving the remaining equations gives, at leading order,
\begin{align}
A_1^{(w)}
&=
\frac{1}{\sqrt{2}}
\exp\left[
\frac{iw\lambda}{2}\ln(2z_0)
\right]
+
\mathcal O(z_0^{-1}),
\label{init-A1}
\\
A_2^{(w)}
&=
-\frac{iw}{\sqrt{2}}
\exp\left[
\frac{iw\lambda}{2}\ln(2z_0)
\right]
+
\mathcal O(z_0^{-1}).
\label{init-A2}
\end{align}
The leading coefficients are independent of $\nu$, as expected: deep inside the horizon, the entropy mass is negligible relative to the physical momentum. Dependence on $\nu$ first appears in corrections of order $z_0^{-1}$. In the strict Bunch--Davies limit, the Kummer-sector coefficients therefore vanish, $B_b^{(w)}\to0$, and the physical solutions receive contributions only from the Tricomi-generated modes.

Retaining the leading Bunch--Davies coefficients, Eq.~\eqref{varphi-F} reduces to
\be
\varphi_b(k,z)
=
\sum_{w=\pm1}
A_b^{(w)}
\mathcal F_U^{(w)}(k,z),
\label{varphi-A}
\ee
with $A_b^{(w)}$ given by Eqs.~\eqref{init-A1} and~\eqref{init-A2}; the corresponding expression for $\sigma_b$ follows by replacing $\mathcal F_U^{(w)}$ with $\mathcal S_U^{(w)}$. Although the individual coefficients retain the arbitrary reference phase associated with $z_0$, they obey
\be
\sum_{b=1}^{2}
A_b^{(w)}A_b^{(w')*}
=
\delta_{ww'}.
\label{AA-cond}
\ee
As we shall see, this unitary relation ensures that the arbitrary reference scale $z_0$ cancels from physical observables.


\section{Power spectrum}
\label{sec:PS}

As an immediate application of the preceding results, we now compute the power spectrum of the curvature perturbation. Using the field redefinition~\eqref{canonical-z}, the dimensionless power spectrum $\Delta_\zeta(k)$ can be expressed in terms of the mode functions $\varphi_b$ as
\be
\Delta_\zeta(k)
=
\frac{k^3}{4\pi^2\epsilon m_{\rm Pl}^2}
\sum_{b=1}^{2}
\left|\varphi_b(k,z)\right|^2.
\label{power-zeta-phi}
\ee
Substituting Eq.~\eqref{varphi-A} into Eq.~\eqref{power-zeta-phi} and using the unitarity relation~\eqref{AA-cond}, we obtain
\be
\Delta_\zeta(k)
=
\frac{k^3}{4\pi^2\epsilon m_{\rm Pl}^2}
\sum_{w=\pm1}
\left|
\mathcal F_U^{(w)}(k,z)
\right|^2.
\ee
Finally, using the superhorizon expansion~\eqref{ZU-small-z-general}, we find, for $\operatorname{Re}\nu<3/2$,
\begin{align}
\Delta_\zeta(k)
&=
\frac{H^2}{8\pi^2\epsilon}
\left|
\frac{
\Gamma\left(\frac34-\frac{\nu}{2}\right)
\Gamma\left(\frac34+\frac{\nu}{2}\right)
}{
\Gamma\left(\frac34-\frac{\nu}{2}+\frac{i\lambda}{2}\right)
\Gamma\left(\frac34+\frac{\nu}{2}+\frac{i\lambda}{2}\right)
}
\right|^2.
\label{power-zeta}
\end{align}
This reproduces the result obtained in Ref.~\cite{Huenupi:2026abj}, and may be analytically continued to complex values of $\nu$.


\section{Schwinger--Keldysh rules}
\label{sec:SK-rules}

In this section, we derive dressed Schwinger--Keldysh rules for computing the correlation functions generated by the cubic interactions considered here, assuming constant $\lambda$ and $\mu/H$. We follow the standard Schwinger--Keldysh formalism reviewed, for example, in Ref.~\cite{Chen:2017ryl}. Since interaction vertices involve time integrals, it is convenient to replace $z$ with conformal time $\tau$. At leading order in slow roll, the two variables are related by
\be
z=-k\tau.
\label{z-to-tau}
\ee
Henceforth, the functions of $z$ introduced in the preceding sections will be regarded as functions of $\tau$ through Eq.~\eqref{z-to-tau}.

\subsection{Wightman functions}

We regard the complete quadratic action, including the mixing proportional to $\lambda$, as the free theory. The interaction-picture fields therefore obey the coupled Eqs.~\eqref{eom-1} and~\eqref{eom-2}, while only the cubic interactions are treated perturbatively. The resulting propagators resum the transfer between curvature and entropy perturbations to all orders in $\lambda$. To construct them, it is convenient to collect the mode functions of the canonically normalized curvature perturbation and the entropy perturbation into the field-space vector
\be
\boldsymbol{\Phi}_b(k,\tau)
\equiv
\begin{pmatrix}
\varphi_b(k,\tau)\\
\sigma_b(k,\tau)
\end{pmatrix}.
\label{def-field-vector}
\ee
We similarly define the basis vectors
\be
\boldsymbol{\Psi}^{(w)}(k,\tau)
\equiv
\begin{pmatrix}
\mathcal F_U^{(w)}(k,\tau)\\
\mathcal S_U^{(w)}(k,\tau)
\end{pmatrix}.
\label{def-Psi-w}
\ee
Writing $\Phi_A(\boldsymbol k,\tau)$ for the Fourier-space field operator, with $A\in\{\varphi,\sigma\}$, we define the Wightman function by
\be
\expval{
\Phi_A(\boldsymbol k,\tau)
\Phi_B(\boldsymbol k',\tau')
}
=
(2\pi)^3
\delta^{(3)}
\left(
\boldsymbol k+\boldsymbol k'
\right)
G^>_{AB}(k;\tau,\tau').
\label{def-Wightman}
\ee
The mode expansion then gives
\be
G^>_{AB}(k;\tau,\tau')
=
\sum_{b=1}^{2}
\Phi_{Ab}(k,\tau)
\Phi_{Bb}^*(k,\tau').
\label{eq-Wightman-matrix}
\ee
Using Eq.~\eqref{varphi-A}, its entropy counterpart, and the unitarity relation~\eqref{AA-cond}, this expression reduces to
\be
G^>_{AB}(k;\tau,\tau')
=
\sum_{w=\pm1}
\Psi_A^{(w)}(k,\tau)
\Psi_B^{(w)*}(k,\tau').
\label{eq-Wightman-Tricomi}
\ee
The dependence on the arbitrary reference time $z_0$ has disappeared after summing over the two independent modes. In components,
\begin{align}
G^>_{\varphi\varphi}(k;\tau,\tau')
&=
\sum_{w=\pm1}
\mathcal F_U^{(w)}(k,\tau)
\mathcal F_U^{(w)*}(k,\tau'),
\label{eq-Gzz}
\\
G^>_{\varphi\sigma}(k;\tau,\tau')
&=
\sum_{w=\pm1}
\mathcal F_U^{(w)}(k,\tau)
\mathcal S_U^{(w)*}(k,\tau'),
\label{eq-Gzs}
\\
G^>_{\sigma\varphi}(k;\tau,\tau')
&=
\sum_{w=\pm1}
\mathcal S_U^{(w)}(k,\tau)
\mathcal F_U^{(w)*}(k,\tau'),
\label{eq-Gsz}
\\
G^>_{\sigma\sigma}(k;\tau,\tau')
&=
\sum_{w=\pm1}
\mathcal S_U^{(w)}(k,\tau)
\mathcal S_U^{(w)*}(k,\tau').
\label{eq-Gss}
\end{align}
Because $\mathcal F_U^{(w)}$ and $\mathcal S_U^{(w)}$ are expressed directly in terms of $f_U^{(w)}$, $g_U^{(w)}$, and $h_U^{(w)}$, the dressed propagators contain no explicit Hankel functions.

\subsection{Bulk-to-bulk propagators}
\label{sub:bulk-to-bulk-prop}

In the Schwinger--Keldysh formalism, each interaction vertex lies on one of the two branches of the time contour, labeled by $a=\pm$. With the Wightman convention introduced above, the four bulk-to-bulk propagators are
\begin{align}
G^{-+}_{AB}(k;\tau,\tau')
&=
G^>_{AB}(k;\tau,\tau'),
\\
G^{+-}_{AB}(k;\tau,\tau')
&=
G^<_{AB}(k;\tau,\tau'),
\\
G^{++}_{AB}(k;\tau,\tau')
&=
\Theta(\tau-\tau')
G^>_{AB}(k;\tau,\tau')
\nn\\
&\quad+
\Theta(\tau'-\tau)
G^<_{AB}(k;\tau,\tau'),
\\
G^{--}_{AB}(k;\tau,\tau')
&=
\Theta(\tau'-\tau)
G^>_{AB}(k;\tau,\tau')
\nn\\
&\quad+
\Theta(\tau-\tau')
G^<_{AB}(k;\tau,\tau'),
\label{eq-SK-propagators-tau}
\end{align}
where the oppositely ordered Wightman function is
\be
G^<_{AB}(k;\tau,\tau')
=
G^>_{BA}(k;\tau',\tau)
=
\left[
G^>_{AB}(k;\tau,\tau')
\right]^*.
\label{eq-lesser-components}
\ee
A line connecting a field $\Phi_A$ on branch $a$ to a field $\Phi_B$ on branch $b$ carries the propagator $G^{ab}_{AB}$. Since these propagators are matrices in field space, a single dressed line may connect a curvature endpoint to an entropy endpoint:
\begin{align}
\begin{tikzpicture}[baseline=-3pt]
  \skbulkminus{v1}{(0,0)}
  \skbulkplus{v2}{(2,0)}
  \skprop{v1}{v2}
  \node[below=1pt,font=\scriptsize] at (1,0) {$k$};
  \node[below=4pt,font=\scriptsize] at (v1) {$\tau,A$};
  \node[below=4pt,font=\scriptsize] at (v2) {$\tau',B$};
\end{tikzpicture}
&=
G^{-+}_{AB}(k;\tau,\tau'),
\\
\begin{tikzpicture}[baseline=-3pt]
  \skbulkplus{v1}{(0,0)}
  \skbulkminus{v2}{(2,0)}
  \skprop{v1}{v2}
  \node[below=1pt,font=\scriptsize] at (1,0) {$k$};
  \node[below=4pt,font=\scriptsize] at (v1) {$\tau,A$};
  \node[below=4pt,font=\scriptsize] at (v2) {$\tau',B$};
\end{tikzpicture}
&=
G^{+-}_{AB}(k;\tau,\tau'),
\\
\begin{tikzpicture}[baseline=-3pt]
  \skbulkplus{v1}{(0,0)}
  \skbulkplus{v2}{(2,0)}
  \skprop{v1}{v2}
  \node[below=1pt,font=\scriptsize] at (1,0) {$k$};
  \node[below=4pt,font=\scriptsize] at (v1) {$\tau,A$};
  \node[below=4pt,font=\scriptsize] at (v2) {$\tau',B$};
\end{tikzpicture}
&=
G^{++}_{AB}(k;\tau,\tau'),
\\
\begin{tikzpicture}[baseline=-3pt]
  \skbulkminus{v1}{(0,0)}
  \skbulkminus{v2}{(2,0)}
  \skprop{v1}{v2}
  \node[below=1pt,font=\scriptsize] at (1,0) {$k$};
  \node[below=4pt,font=\scriptsize] at (v1) {$\tau,A$};
  \node[below=4pt,font=\scriptsize] at (v2) {$\tau',B$};
\end{tikzpicture}
&=
G^{--}_{AB}(k;\tau,\tau').
\end{align}
These propagators already contain the effects of the quadratic mixing to all orders in $\lambda$, so no separate quadratic transfer vertices are required.

\subsection{Bulk-to-boundary propagators}
\label{sub:bulk-to-boundary-prop}

The external observable is the physical curvature perturbation $\zeta= \varphi / \sqrt{2\epsilon} m_{\rm Pl}$. For an external curvature perturbation evaluated at $\tau_f$, we therefore define
\be
K_A^r(k;\tau_f,\tau)
\equiv
\frac{1}{\sqrt{2\epsilon} m_{\rm Pl}}\,
G^{+r}_{\varphi A}(k;\tau_f,\tau),
\qquad
r=\pm,
\label{def-bulk-boundary}
\ee
where $A\in\{\varphi,\sigma\}$. Since $\tau_f$ is later than every interaction vertex, no Heaviside functions are needed:
\begin{align}
K_A^+(k;\tau_f,\tau)
&=
\frac{1}{\sqrt{2\epsilon} m_{\rm Pl}}\,
G^>_{\varphi A}(k;\tau_f,\tau),
\label{eq-Kplus}
\\
K_A^-(k;\tau_f,\tau)
&=
\frac{1}{\sqrt{2\epsilon} m_{\rm Pl}}\,
G^<_{\varphi A}(k;\tau_f,\tau).
\label{eq-Kminus}
\end{align}
The two $+$-branch bulk-to-boundary propagators entering the cubic correlators are therefore
\begin{align}
K_\varphi^+(k;\tau_f,\tau)
&=
\frac{1}{\sqrt{2\epsilon} m_{\rm Pl}}
\sum_{w=\pm1}
\mathcal F_U^{(w)}(k,\tau_f)
\mathcal F_U^{(w)*}(k,\tau),
\label{eq-Kzeta}
\\
K_\sigma^+(k;\tau_f,\tau)
&=
\frac{1}{\sqrt{2\epsilon} m_{\rm Pl}}
\sum_{w=\pm1}
\mathcal F_U^{(w)}(k,\tau_f)
\mathcal S_U^{(w)*}(k,\tau).
\label{eq-Ksigma}
\end{align}
The corresponding propagators on the opposite branch are their complex conjugates:
\be
K_A^-(k;\tau_f,\tau)
=
\left[
K_A^+(k;\tau_f,\tau)
\right]^*.
\ee
The diagrammatic assignments for a boundary curvature perturbation connected to a bulk field $\Phi_A$ are
\begin{align}
\begin{tikzpicture}[baseline=-3pt]
  \skboundary{v1}{(0,0)}
  \skbulkplus{v2}{(2,0)}
  \skprop{v1}{v2}
  \node[below=1pt,font=\scriptsize] at (1,0) {$k$};
  \node[below=4pt,font=\scriptsize] at (v1) {$\zeta$};
  \node[below=4pt,font=\scriptsize] at (v2) {$\tau,A$};
\end{tikzpicture}
&=
K_A^+(k;\tau_f,\tau),
\\
\begin{tikzpicture}[baseline=-3pt]
  \skboundary{v1}{(0,0)}
  \skbulkminus{v2}{(2,0)}
  \skprop{v1}{v2}
  \node[below=1pt,font=\scriptsize] at (1,0) {$k$};
  \node[below=4pt,font=\scriptsize] at (v1) {$\zeta$};
  \node[below=4pt,font=\scriptsize] at (v2) {$\tau,A$};
\end{tikzpicture}
&=
K_A^-(k;\tau_f,\tau).
\end{align}

\subsection{Covariant derivatives on propagators}
\label{sub:cov-endpoint}

The cubic interactions contain the covariant derivative $D_t$ defined in Eq.~\eqref{cov-D_t}. In conformal time, the corresponding derivative is
\be
D_\tau\varphi
\equiv
\partial_\tau\varphi
-
a(\tau)\lambda H\sigma.
\label{def-Dtau-zeta}
\ee
Thus, $D_\tau\varphi=aD_t\varphi$. Using $z=-k\tau$ and $aH=-1/\tau$ at the order considered here, this derivative is related to $D_z$ in Eq.~\eqref{cov-D_z} by
\be
D_\tau\varphi
=
-kD_z\varphi.
\label{eq-Dtau-Dz}
\ee
We define the differentiated bulk-to-boundary propagator by
\be
K_{\mathcal D}^\pm(k;\tau_f,\tau)
\equiv
\partial_\tau K_\varphi^{\pm}(k;\tau_f,\tau)
-
a(\tau)\lambda H
K_\sigma^\pm(k;\tau_f,\tau) .
\label{def-KD-tau}
\ee
Using the explicit mode-function representation, its $+$-branch component becomes
\be
K_{\mathcal D}^+(k;\tau_f,\tau)
=
\frac{-k}{\sqrt{2\epsilon} m_{\rm Pl}} \!\!
\sum_{w=\pm1} \!\!
\mathcal F_U^{(w)}(k,\tau_f) \!
\left[ \!
D_z\mathcal F_U^{(w)} \! (k,z)
\right]^* \!\!\! ,
\label{eq-KD-plus-compact}
\ee
where $D_z\mathcal F_U^{(w)}$ is given in Eq.~\eqref{eq-Dz-Z}. The corresponding propagator on the opposite branch is
\be
K_{\mathcal D}^-(k;\tau_f,\tau)
=
\left[
K_{\mathcal D}^+(k;\tau_f,\tau)
\right]^*.
\label{eq-KD-minus-compact}
\ee

Bulk-to-bulk propagators with one differentiated endpoint are defined analogously. Writing $z=-k\tau$ and $z'=-k\tau'$, the differentiated Wightman functions are
\begin{align}
G^>_{\mathcal D B}(k;\tau,\tau')
&=
-k
\sum_{w=\pm1}
\left[
D_z\mathcal F_U^{(w)}(k,z)
\right]
\Psi_B^{(w)*}(k,\tau'),
\label{eq-GD-B-compact}
\\
G^>_{A\mathcal D}(k;\tau,\tau')
&=
-k
\sum_{w=\pm1}
\Psi_A^{(w)}(k,\tau)
\left[
D_{z'}\mathcal F_U^{(w)}(k,z')
\right]^*.
\label{eq-GA-D-compact}
\end{align}
Here $A,B\in\{\varphi,\sigma\}$, and the label $\mathcal D$ denotes an endpoint carrying the operator $D_\tau\varphi$. If both endpoints are differentiated, one instead obtains
\be
G^>_{\mathcal D\mathcal D}(k;\tau,\tau')
=
k^2
\sum_{w=\pm1}
\left[
D_z\mathcal F_U^{(w)}(k,z)
\right]
\left[
D_{z'}\mathcal F_U^{(w)}(k,z')
\right]^*.
\label{eq-GDD-compact}
\ee
The oppositely ordered functions follow by reversing the endpoint ordering:
\begin{align}
G^<_{\mathcal D B}(k;\tau,\tau')
&=
G^>_{B\mathcal D}(k;\tau',\tau),
\\
G^<_{A\mathcal D}(k;\tau,\tau')
&=
G^>_{\mathcal D A}(k;\tau',\tau),
\\
G^<_{\mathcal D\mathcal D}(k;\tau,\tau')
&=
G^>_{\mathcal D\mathcal D}(k;\tau',\tau).
\end{align}
The associated Schwinger--Keldysh propagators are then constructed using the same time-ordering rules as in Eq.~\eqref{eq-SK-propagators-tau}, with one or both field labels replaced by $\mathcal D$. For example,
\begin{align}
G^{++}_{\mathcal D B}(k;\tau,\tau')
&=
\Theta(\tau-\tau')
G^>_{\mathcal D B}(k;\tau,\tau')
\nn\\
&\quad+
\Theta(\tau'-\tau)
G^<_{\mathcal D B}(k;\tau,\tau'),
\\
G^{--}_{\mathcal D B}(k;\tau,\tau')
&=
\Theta(\tau'-\tau)
G^>_{\mathcal D B}(k;\tau,\tau')
\nn\\
&\quad+
\Theta(\tau-\tau')
G^<_{\mathcal D B}(k;\tau,\tau').
\end{align}
For separated bulk vertices, this construction is equivalent to acting with the covariant derivative on the corresponding endpoint of the dressed propagator. Possible coincident-time contact terms must instead be treated through the interaction Hamiltonian and should not be inferred by differentiating the Heaviside functions.

\subsection{Vertices}
\label{sub:vertices}

Expressing Eq.~\eqref{S3-covariant} in terms of the canonically normalized field $\varphi=\sqrt{2\epsilon}\, m_{\rm Pl}\zeta$ and using $\dd{\tau}=\dd{t}/a$ gives
\begin{align}
S_{\rm int}^{(3)}
= & 
\int\dd{\tau}\dd[3]{x} \frac{1}{\sqrt{2 \epsilon}\, m_{\rm Pl}}
\bigg[
\frac{a^2 \alpha}{2!}\sigma
\left(
D_\tau\varphi
\right)^2 \nn 
\\
&\quad \qquad +
\frac{a^3\beta H }{2!}\sigma^2D_\tau\varphi
+
\frac{a^4 \gamma H^2}{3!}\sigma^3
\bigg],
\label{eq-cubic-action-tau}
\end{align}
where $D_\tau\varphi$ is defined in Eq.~\eqref{def-Dtau-zeta}. At the cubic order relevant for the tree-level bispectra considered below, the interaction Hamiltonian is obtained as $H_{\rm int}^{(3)}=-L_{\rm int}^{(3)}$ after expressing the velocities in terms of the free canonical variables. The factorials in Eq.~\eqref{eq-cubic-action-tau} are canceled by permutations of the identical legs, yielding the vertex coefficients displayed below.

The interaction $\sigma(D_t\varphi)^2$, proportional to $\alpha$, produces two differentiated curvature endpoints. Its assignments on the two contour branches are
\begin{align}
\begin{tikzpicture}[baseline=-3pt]
  \skbulkplus{v}{(0,0)}
  \draw[sk/prop] (v) -- +(-1,0);
  \draw[sk/prop] (v) -- +(0.5,+0.7);
  \draw[sk/prop] (v) -- +(0.5,-0.7);
  \node[below=4pt] at (-0.2,0) {$\alpha$};
  \node[above=1pt,font=\scriptsize] at (0.5,+0.7) {$D_\tau\varphi,\k_2$};
  \node[below=1pt,font=\scriptsize] at (0.5,-0.7) {$D_\tau\varphi,\k_3$};
  \node[above=1pt,font=\scriptsize] at (-1,0) {$\sigma,\k_1$};
\end{tikzpicture}
\!\!\!\!\!\!\!\!\!\!\!
&=
+\frac{i \alpha(2\pi)^3}{\sqrt{2 \epsilon} \, m_{\rm Pl}}\delta^{(3)}(\boldsymbol K) \!\!
\int_{-\infty(1-i0^+)}^{\tau_f}
\hspace{-35pt}\dd{\tau} 
a^2 \,D_{\tau,2}D_{\tau,3},
\\
\begin{tikzpicture}[baseline=-3pt]
  \skbulkminus{v}{(0,0)}
  \draw[sk/prop] (v) -- +(-1,0);
  \draw[sk/prop] (v) -- +(0.5,+0.7);
  \draw[sk/prop] (v) -- +(0.5,-0.7);
  \node[below=4pt] at (-0.2,0) {$\alpha$};
  \node[above=1pt,font=\scriptsize] at (0.5,+0.7) {$D_\tau\varphi,\k_2$};
  \node[below=1pt,font=\scriptsize] at (0.5,-0.7) {$D_\tau\varphi,\k_3$};
  \node[above=1pt,font=\scriptsize] at (-1,0) {$\sigma,\k_1$};
\end{tikzpicture}
\!\!\!\!\!\!\!\!\!\!\!
&=
- \frac{i \alpha(2\pi)^3}{\sqrt{2 \epsilon}  \, m_{\rm Pl}}\delta^{(3)}(\boldsymbol K) \!\!
\int_{-\infty(1+i0^+)}^{\tau_f}
\hspace{-35pt}\dd{\tau} 
a^2 \,D_{\tau,2}D_{\tau,3},
\end{align}
where $\boldsymbol K=\boldsymbol k_1+\boldsymbol k_2+\boldsymbol k_3$. The conjugate $i\epsilon$ prescriptions select the interacting vacuum on the two branches of the contour. The operators $D_{\tau,2}$ and $D_{\tau,3}$ act on the propagators carrying momenta $k_2$ and $k_3$, respectively. Similarly, the interaction $\sigma^2D_t\varphi$, proportional to $\beta$, produces one differentiated curvature endpoint:
\begin{align}
\begin{tikzpicture}[baseline=-3pt]
  \skbulkplus{v}{(0,0)}
  \draw[sk/prop] (v) -- +(-1,0);
  \draw[sk/prop] (v) -- +(0.5,+0.7);
  \draw[sk/prop] (v) -- +(0.5,-0.7);
  \node[below=4pt] at (-0.2,0) {$\beta$};
  \node[above=1pt,font=\scriptsize] at (0.5,+0.7) {$\sigma,\k_2$};
  \node[below=1pt,font=\scriptsize] at (0.5,-0.7) {$\sigma,\k_3$};
  \node[above=1pt,font=\scriptsize] at (-1,0) {$D_\tau\varphi,\k_1$};
\end{tikzpicture}
\!\!\!\!\!\!\!\!\!
&=
+\frac{i \beta (2\pi)^3}{\sqrt{2 \epsilon} m_{\rm Pl} }  \delta^{(3)}(\boldsymbol K)
\int_{-\infty(1-i0^+)}^{\tau_f}
\hspace{-35pt}\dd{\tau} 
a^3 \,D_{\tau,1},
\\
\begin{tikzpicture}[baseline=-3pt]
  \skbulkminus{v}{(0,0)}
  \draw[sk/prop] (v) -- +(-1,0);
  \draw[sk/prop] (v) -- +(0.5,+0.7);
  \draw[sk/prop] (v) -- +(0.5,-0.7);
  \node[below=4pt] at (-0.2,0) {$\beta$};
  \node[above=1pt,font=\scriptsize] at (0.5,+0.7) {$\sigma,\k_2$};
  \node[below=1pt,font=\scriptsize] at (0.5,-0.7) {$\sigma,\k_3$};
  \node[above=1pt,font=\scriptsize] at (-1,0) {$D_\tau\varphi,\k_1$};
\end{tikzpicture}
\!\!\!\!\!\!\!\!\!
&=
-\frac{i \beta(2\pi)^3}{\sqrt{2 \epsilon} m_{\rm Pl} }   \delta^{(3)}(\boldsymbol K)
\int_{-\infty(1+i0^+)}^{\tau_f}
\hspace{-35pt}\dd{\tau}
a^3 \,D_{\tau,1}.
\end{align}
Here $D_{\tau,1}$ acts on the propagator carrying momentum $k_1$. Finally, the interaction $\sigma^3$ gives
\begin{align}
\begin{tikzpicture}[baseline=-3pt]
  \skbulkplus{v}{(0,0)}
  \draw[sk/prop] (v) -- +(-1,0);
  \draw[sk/prop] (v) -- +(0.5,+0.7);
  \draw[sk/prop] (v) -- +(0.5,-0.7);
  \node[below=4pt] at (-0.2,0) {$\gamma$};
  \node[above=1pt,font=\scriptsize] at (0.5,+0.7) {$\sigma,\k_2$};
  \node[below=1pt,font=\scriptsize] at (0.5,-0.7) {$\sigma,\k_3$};
  \node[above=1pt,font=\scriptsize] at (-1,0) {$\sigma,\k_1$};
\end{tikzpicture}
\!\!\!\!\!\!\!\!\!
&=
+\frac{i \gamma (2\pi)^3}{\sqrt{2 \epsilon} m_{\rm Pl}}  \delta^{(3)}(\boldsymbol K)
\int_{-\infty(1-i0^+)}^{\tau_f}
\hspace{-35pt}\dd{\tau} a^4 ,
\\
\begin{tikzpicture}[baseline=-3pt]
  \skbulkminus{v}{(0,0)}
  \draw[sk/prop] (v) -- +(-1,0);
  \draw[sk/prop] (v) -- +(0.5,+0.7);
  \draw[sk/prop] (v) -- +(0.5,-0.7);
  \node[below=4pt] at (-0.2,0) {$\gamma$};
  \node[above=1pt,font=\scriptsize] at (0.5,+0.7) {$\sigma,\k_2$};
  \node[below=1pt,font=\scriptsize] at (0.5,-0.7) {$\sigma,\k_3$};
  \node[above=1pt,font=\scriptsize] at (-1,0) {$\sigma,\k_1$};
\end{tikzpicture}
\!\!\!\!\!\!\!\!\!
&=
-\frac{i \gamma (2\pi)^3}{\sqrt{2 \epsilon} m_{\rm Pl} }  \delta^{(3)}(\boldsymbol K)
\int_{-\infty(1+i0^+)}^{\tau_f}
\hspace{-35pt}\dd{\tau} a^4  .
\end{align}

\subsection{Diagrammatic rules}

We conclude by summarizing the rules for computing correlation functions of $\zeta$. An $n$-point function is obtained by summing all diagrams assembled from the dressed bulk-to-bulk propagators, bulk-to-boundary propagators, and cubic vertices introduced above. Each bulk vertex is assigned to one of the two Schwinger--Keldysh branches, and every endpoint carries a field label in the basis $(\varphi,\sigma)$. A line joining two bulk endpoints carries the corresponding bulk-to-bulk propagator, whereas a line ending on the future boundary carries the bulk-to-boundary propagator associated with an external $\zeta$.

If an endpoint contains $D_\tau\varphi$, the ordinary curvature endpoint is replaced by the differentiated endpoint defined in Subsection~\ref{sub:cov-endpoint}. This prescription applies to both bulk-to-boundary and bulk-to-bulk propagators. After inserting the vertex assignments of Subsection~\ref{sub:vertices}, one integrates over every bulk time along the appropriate tilted contour and over every internal momentum with measure $(2\pi)^{-3}\dd[3]{q}$. The product of the momentum-conserving delta functions at the vertices leaves a single overall Dirac delta enforcing conservation of the external momenta. For the tree-level contact diagrams considered below, the two complete branch contributions are complex conjugates. Their sum is therefore twice the real part of the $+$-branch diagram or, equivalently, minus twice the imaginary part of the corresponding time integral after the explicit $+i$ vertex factor has been removed.


\section{Bispectrum}

For each of the three cubic interactions, the three-point function receives one contribution from each branch of the Schwinger--Keldysh contour:
\begin{align}
\expval{
\zeta_{\boldsymbol k_1}
\zeta_{\boldsymbol k_2}
\zeta_{\boldsymbol k_3}
} (\tau_f)
&=
\begin{tikzpicture}[baseline=-3pt]
  \draw[sk/prop] (-1.25,0) -- (1.25,0);
  \node[left,font=\scriptsize] at (-1.25,0) {$\tau_f$};
  \skbulkplus{v}{(0,-1)}
  \skboundary{v1}{(-1,0)}
  \skboundary{v2}{(0,0)}
  \skboundary{v3}{(1,0)}
  \draw[sk/prop] (v) -- (v1);
  \draw[sk/prop] (v) -- (v2);
  \draw[sk/prop] (v) -- (v3);
  \node[above=5pt,font=\scriptsize] at (v1) {$\k_1$};
  \node[above=5pt,font=\scriptsize] at (v2) {$\k_2$};
  \node[above=5pt,font=\scriptsize] at (v3) {$\k_3$};
\end{tikzpicture}
\nn\\
&\quad+
\begin{tikzpicture}[baseline=-3pt]
  \draw[sk/prop] (-1.25,0) -- (1.25,0);
  \node[left,font=\scriptsize] at (-1.25,0) {$\tau_f$};
  \skbulkminus{v}{(0,-1)}
  \skboundary{v1}{(-1,0)}
  \skboundary{v2}{(0,0)}
  \skboundary{v3}{(1,0)}
  \draw[sk/prop] (v) -- (v1);
  \draw[sk/prop] (v) -- (v2);
  \draw[sk/prop] (v) -- (v3);
  \node[above=5pt,font=\scriptsize] at (v1) {$\k_1$};
  \node[above=5pt,font=\scriptsize] at (v2) {$\k_2$};
  \node[above=5pt,font=\scriptsize] at (v3) {$\k_3$};
\end{tikzpicture}.
\end{align}
The bispectrum $B_\zeta(k_1,k_2,k_3)$ is defined by
\be
\expval{
\zeta_{\boldsymbol k_1} 
\zeta_{\boldsymbol k_2} 
\zeta_{\boldsymbol k_3} 
} (\tau_f)
\equiv
(2\pi)^3
\delta^{(3)}
\left(
\boldsymbol K
\right)
B_\zeta(k_1,k_2,k_3),
\label{def-bispectrum}
\ee
where $\boldsymbol K=\boldsymbol k_1+\boldsymbol k_2+\boldsymbol k_3$ denotes the total momentum vector, while $K\equiv k_1+k_2+k_3$ denotes the sum of its magnitudes. We now derive the contribution generated by each cubic interaction. 

We express the results in terms of the scale-invariant dimensionless shape function $\mathcal S(k_1,k_2,k_3)$, defined via: 
\begin{equation}
B_\zeta(k_1,k_2,k_3) \equiv \frac{9}{10} \frac{
(2\pi)^4\Delta_\zeta^2
}{(k_1k_2k_3)^2} \sum_i f^{(i)}_{\rm NL} \mathcal S_i(k_1,k_2,k_3) ,
\end{equation}
where $\Delta_\zeta$ is the dimensionless power spectrum in Eq.~\eqref{power-zeta}, and the index $i$-labels the interaction inducing the shape $\mathcal S_i$. The parameter $f^{(i)}_{\rm NL}$ is fixed by the normalization condition $\mathcal S_i (k,k,k) = 1$.

\subsection{$\alpha$ interaction}

\begin{figure*}[t!]
\centering
\includegraphics[width=\textwidth]
    {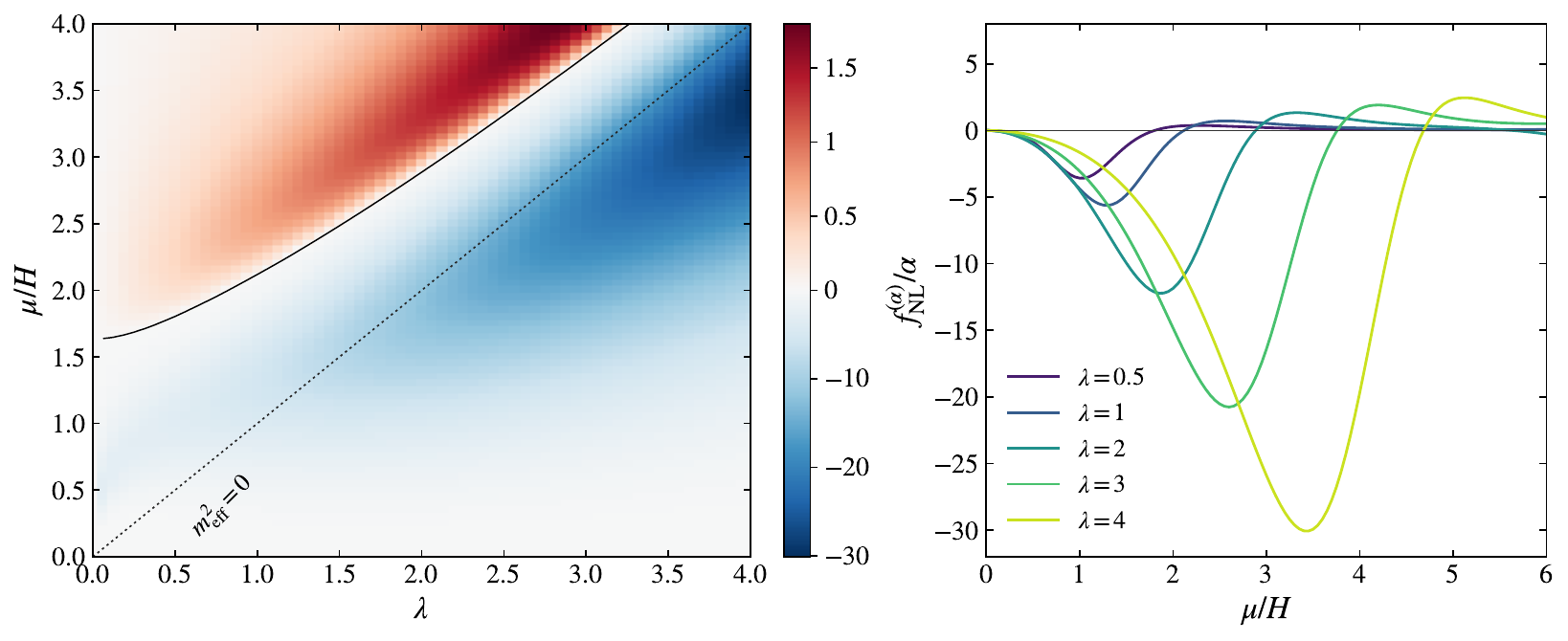}
    \caption{
        The normalized equilateral amplitude
        $f_{\rm NL}^{(\alpha)} / \alpha$  as a function of $\lambda$ and $\mu/H$. The left panel shows the full parameter plane, while the right panel displays representative slices for fixed values of $\lambda$.
    }
    \label{fig:equilateral_alpha}
\end{figure*}

We start by considering the computation of the bispectrum induced by the operator $\sigma(D_t\varphi)^2$. According to the Schwinger-Keldysh rules discussed in the previous section, the tree-level bispectrum $B_{\alpha}$ proportional to the coupling $\alpha$ is:
\begin{align}
B_\alpha(k_1,k_2,k_3)
&=
\frac{-2\alpha}{\sqrt{2\epsilon} m_{\rm Pl}}
\Im \!
\int_{-\infty(1-i0^+)}^{\tau_f} \!\!\!\!\!\!\!\!\!\!\!\!\!\!\!\!\!\!\!\!\!
\dd{\tau}\,
a^2(\tau) \,
K_{\mathcal D}^+
\left(
k_1;\tau_f,\tau
\right)
\nn\\
&\quad\times
K_{\mathcal D}^+
\left(
k_2;\tau_f,\tau
\right)
K_\sigma^+
\left(
k_3;\tau_f,\tau
\right)
\nn\\
&\quad+
2\ \text{cyclic permutations}.
\label{eq-B-alpha-tau}
\end{align}
For later convenience, we take the complex conjugate of the time integral, which reverses the sign of its imaginary part and replaces the $+$ contour by its conjugate. Using the explicit dressed bulk-to-boundary propagators then gives
\begin{align}
B_\alpha
&=
\frac{2\alpha}{(2\epsilon)^2 m_{\rm Pl}^4}
\Im
\int_{-\infty(1+i0^+)}^{\tau_f} \!\!\!\!\!\!\!\!\!\!\!\!\!\!\!\!\!\!\!\!\!
\dd{\tau}\,
a^2(\tau)\,k_1k_2 \!\!\!\!\!\!\!\!\!
\sum_{w_1,w_2,w_3=\pm1}
\nn\\
&\quad\times
\mathcal F_U^{(w_1)*}(k_1,\tau_f)
\mathcal F_U^{(w_2)*}(k_2,\tau_f)
\mathcal F_U^{(w_3)*}(k_3,\tau_f)
\nn\\
&\quad\times
\left[
D_{z_1}\mathcal F_U^{(w_1)}(k_1,z_1)
\right]
\left[
D_{z_2}\mathcal F_U^{(w_2)}(k_2,z_2)
\right]
\nn\\
& \quad\times \mathcal S_U^{(w_3)}(k_3,\tau)+
2\ \text{cyclic permutations},
\label{eq-B-alpha-modes}
\end{align}
where $z_i=-k_i\tau$. This expression is exact in the mixing strength $\lambda$ and the entropy-mass parameter $\mu/H$, while remaining perturbative in the cubic coupling $\alpha$. All dependence on the quadratic curvature--entropy transfer is encoded in the dressed functions $D_z\mathcal F_U^{(w)}$ and $\mathcal S_U^{(w)}$. Using Eqs.~\eqref{eq-S-T} and~\eqref{eq-Dz-Z}, the nontrivial time dependence can be isolated in the reduced vertex integral
\begin{align}
I_\alpha^{(\boldsymbol w)}(k_1,k_2,k_3)
&=
\int_{-\infty(1+i0^+)}^{\tau_f} \!\!\!\!\!\!\!\!\!\!\!\!\!\!\!\!\!\!\!\!\!
\dd{\tau}\,
\tau^2e^{-iK\tau}
\nn\\
& \!\!\!\!\!\!\!\!\!\!\!\!\!\!\!\!\!\!\!\! \times
g_U^{(w_1)}(-k_1\tau)
g_U^{(w_2)}(-k_2\tau)
h_U^{(w_3)}(-k_3\tau).
\label{I-alpha-unstarred}
\end{align}
The integral representations~\eqref{g-int-2} and~\eqref{h-int-2} reduce the time integral to
\be
\int_{-\infty(1+i0^+)}^{\tau_f} \!\!\!\!\!\!\!\!\!\!\!\!\!\!\!\!\!\!\!\!\!
\dd{\tau}\,
\tau^2e^{-i\Omega\tau}
=
-i
\left[
\frac{2}{\Omega^3}
+
\frac{2i\tau_f}{\Omega^2}
-
\frac{\tau_f^2}{\Omega}
\right]
e^{-i\tau_f\Omega},
\label{tau-int-Omega}
\ee
where
\be
\Omega(\boldsymbol t)
\equiv
K+2\left(
k_1t_1+k_2t_2+k_3t_3
\right).
\ee
Substitution into Eq.~\eqref{I-alpha-unstarred} yields
\begin{align}
I_\alpha^{(\boldsymbol w)}
&=
-i
\frac{
e^{\pi(w_1+w_2+w_3)\lambda/4}
}{
2\sqrt{2}\,
\Gamma(a_1)
\Gamma(a_2)
\Gamma(1+a_3)
}
\nn\\
&\quad\times
\int_0^\infty\dd{t_1}
\int_0^\infty\dd{t_2}
\int_0^\infty\dd{t_3}
\left[
\frac{2}{\Omega^3}
+
\frac{2i\tau_f}{\Omega^2}
-
\frac{\tau_f^2}{\Omega}
\right]
\nn\\
&\quad\times
e^{-i\tau_f\Omega}
D_\nu^{(w_1)}(-t_1)
D_\nu^{(w_2)}(-t_2)
D_\nu^{(w_3)}(-t_3)
\nn\\
&\quad\times
\frac{1}{t_1^2}
\left(
\frac{t_1}{1+t_1}
\right)^{1+a_1}
\frac{1}{t_2^2}
\left(
\frac{t_2}{1+t_2}
\right)^{1+a_2}
\left(
\frac{t_3}{1+t_3}
\right)^{a_3} \!\!\! ,
\label{Ialpha-result}
\end{align}
where we have defined:
\begin{align}
a_i\equiv & \frac{iw_i\lambda}{2}, \\
\boldsymbol w \equiv & (w_1,w_2,w_3).
\label{def-ai}
\end{align}
Equation~\eqref{Ialpha-result} is well suited both to numerical evaluation and to an analytic study of the parameter dependence. Figure~\ref{fig:equilateral_alpha}, for example, shows the normalized equilateral amplitude $f_{\rm NL}^{(\alpha)}/\alpha$ as a function of $\lambda$ and $\mu/H$. Since $f_{\rm NL}^{(\alpha)}$ is normalized by the exact curvature power spectrum, the ratio incorporates curvature--isocurvature transfer nonperturbatively. The left panel shows the full parameter plane, with the solid black curve denoting $f_{\rm NL}^{(\alpha)}/\alpha=0$ and the dotted diagonal marking $m_{\rm eff}^{2}=\mu^{2}-H^{2}\lambda^{2}=0$. The right panel shows fixed-$\lambda$ slices.

\begin{figure*}[t!]
\centering
\includegraphics[width=\textwidth]
    {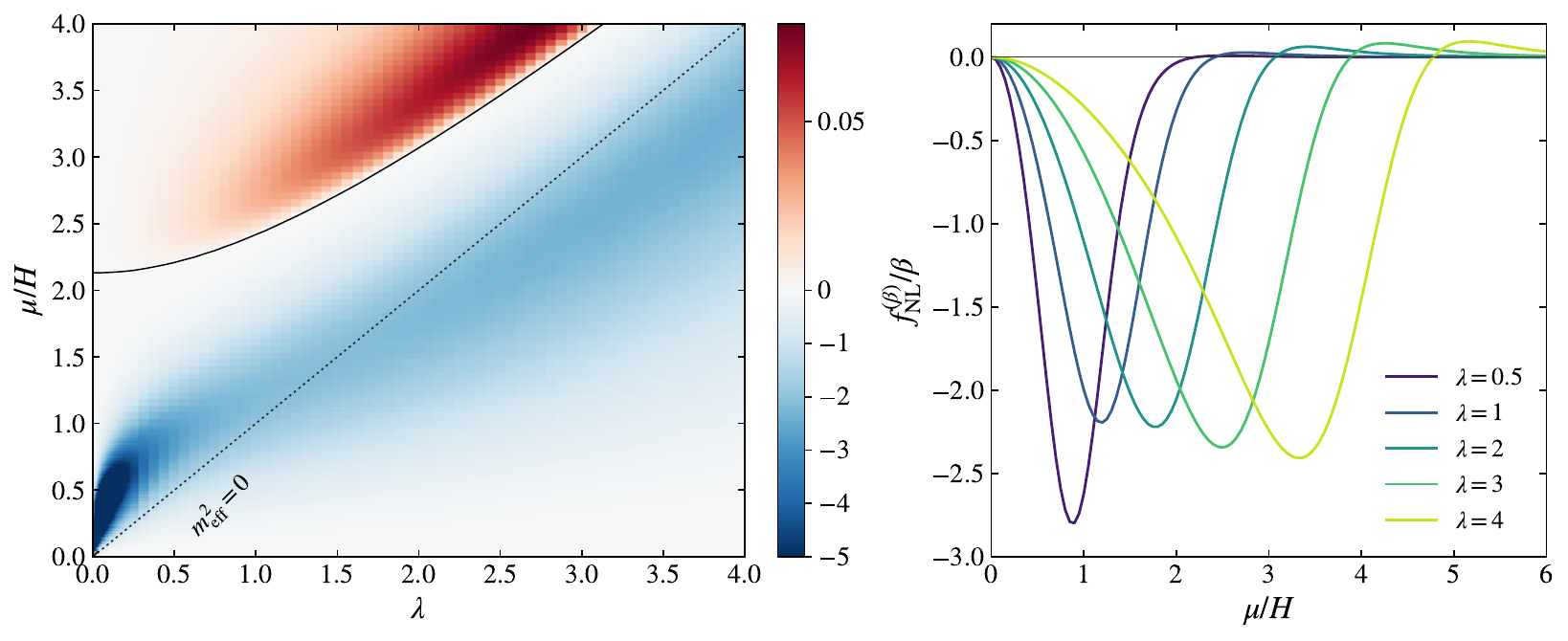}
    \caption{
        Normalized equilateral amplitude
        $f_{\rm NL}^{(\beta)} / \beta$ as a function of $\lambda$ and $\mu/H$.
        The left panel shows the full parameter plane, while the right panel
        displays representative fixed-$\lambda$ slices.
    }
    \label{fig:equilateral_beta}
\end{figure*}

\subsection{$\beta$ interaction}

Let us now examine the bispectrum induced by the second operator $\sigma^2D_t\varphi$. In this case, the Schwinger-Keldysh rules give
\begin{align}
B_\beta(k_1,k_2,k_3)
&=
\frac{-2\beta}{\sqrt{2\epsilon} m_{\rm Pl}}
\Im \!
\int_{-\infty(1-i0^+)}^{\tau_f} \!\!\!\!\!\!\!\!\!\!\!\!\!\!\!\!\!\!\!\!\!
\dd{\tau}\,
a^3(\tau) \,
K_{\mathcal D}^+
\left(
k_1;\tau_f,\tau
\right)
\nn\\
&\quad\times
K_\sigma^+
\left(
k_2;\tau_f,\tau
\right)
K_\sigma^+
\left(
k_3;\tau_f,\tau
\right)
\nn\\
&\quad+
2\ \text{cyclic permutations}.
\label{eq-B-beta-tau}
\end{align}
Taking the complex conjugate of the time integral and inserting the dressed propagators gives
\begin{align}
B_\beta
&=
-\frac{2\beta H}{(2\epsilon)^2 m_{\rm Pl}^4}
\Im
\int_{-\infty(1+i0^+)}^{\tau_f} \!\!\!\!\!\!\!\!\!\!\!\!\!\!\!\!\!\!\!\!\!
\dd{\tau}\,
a^3(\tau) \,k_1 \!\!\!\!\!\!\!\!\!
\sum_{w_1,w_2,w_3=\pm1}
\nn\\
&\quad\times
\mathcal F_U^{(w_1)*}(k_1,\tau_f)
\mathcal F_U^{(w_2)*}(k_2,\tau_f)
\mathcal F_U^{(w_3)*}(k_3,\tau_f)
\nn\\
&\quad\times
\left[
D_{z_1}\mathcal F_U^{(w_1)}(k_1,z_1)
\right]
\mathcal S_U^{(w_2)}(k_2,\tau)
\mathcal S_U^{(w_3)}(k_3,\tau)
\nn\\
&\quad+
2\ \text{cyclic permutations},
\label{eq-B-beta-modes}
\end{align}
where $z_1=-k_1\tau$. The reduced vertex integral is
\begin{align}
I_\beta^{(\boldsymbol w)}(k_1,k_2,k_3)
&=
\int_{-\infty(1+i0^+)}^{\tau_f} \!\!\!\!\!\!\!\!\!\!\!\!\!\!\!\!\!\!\!\!\!
\dd{\tau}\,
\tau^2e^{-iK\tau}
\nn\\
&\!\!\!\!\!\!\!\!\!\!\!\!\!\!\!\!\!\!\!\!\! \times
g_U^{(w_1)}(-k_1\tau)
h_U^{(w_2)}(-k_2\tau)
h_U^{(w_3)}(-k_3\tau).
\end{align}
Using Eqs.~\eqref{g-int-2}, \eqref{h-int-2}, and~\eqref{tau-int-Omega}, one obtains
\begin{align}
I_\beta^{(\boldsymbol w)}
&=
-i
\frac{
e^{\pi(w_1+w_2+w_3)\lambda/4}
}{
2\sqrt{2}\,
\Gamma(a_1)
\Gamma(1+a_2)
\Gamma(1+a_3)
}
\nn\\
&\quad\times
\int_0^\infty\dd{t_1}
\int_0^\infty\dd{t_2}
\int_0^\infty\dd{t_3}
\left[
\frac{2}{\Omega^3}
+
\frac{2i\tau_f}{\Omega^2}
-
\frac{\tau_f^2}{\Omega}
\right]
\nn\\
&\quad\times
e^{-i\tau_f\Omega}
D_\nu^{(w_1)}(-t_1)
D_\nu^{(w_2)}(-t_2)
D_\nu^{(w_3)}(-t_3)
\nn\\
&\quad\times
\frac{1}{t_1^2}
\left(
\frac{t_1}{1+t_1}
\right)^{1+a_1}
\left(
\frac{t_2}{1+t_2}
\right)^{a_2}
\left(
\frac{t_3}{1+t_3}
\right)^{a_3}.
\label{Ibeta-result}
\end{align}

As in the previous case, we may use this result to extract information about the dependence of the bispectrum on the parameter space spanned by $\mu$ and $\lambda$. Figure~\ref{fig:equilateral_beta} shows the normalized equilateral amplitude $f_{\rm NL}^{(\beta)}/\beta$ as a function of $\lambda$ and $\mu$. The left panel shows the full parameter plane, with the solid black curve denoting $f_{\rm NL}^{(\beta)}/\beta=0$ and the dotted diagonal marking $m_{\rm eff}^{2}=\mu^{2}-H^{2}\lambda^{2}=0$. The right panel shows fixed-$\lambda$ slices. The behavior near $(\lambda,\mu)=(0,0)$ is nonuniform. More to the point, defining $\Delta \equiv\frac32-\nu \simeq\frac{\mu^2}{3H^2}$, near the origin one finds
\begin{equation}
\frac{f_{\rm NL}^{(\beta)}}{\beta}
\simeq
-\frac{3\lambda^2\Delta}
{\left(\lambda^2+\Delta^2\right)^2}.
\label{eq:beta-origin}
\end{equation}
Thus, at fixed $\lambda>0$, the amplitude vanishes as $\mu/H\to0$, whereas along a trajectory respecting $\lambda=c\,\mu/H$, for some positive parameter $c$, it approaches the value $-1/c^2$. Along other trajectories the double limit can be unbounded. The feature near the origin thus reflects the noncommutativity of the massless, zero-mixing, and late-time limits, as discussed in~\cite{Huenupi:2026abj}.

\subsection{$\gamma$ interaction}

\begin{figure*}[t!]
\centering
\includegraphics[width=\textwidth]
    {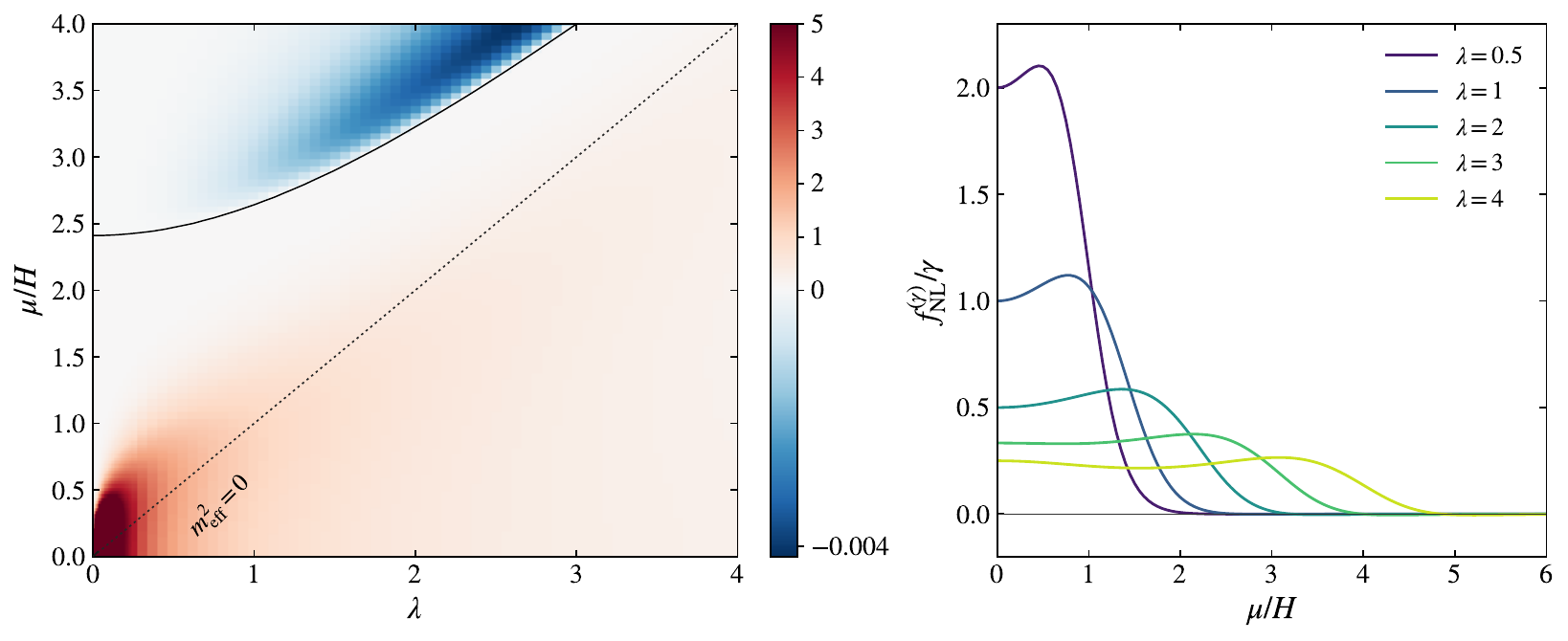}
    \caption{
        Normalized equilateral amplitude
        $f_{\rm NL}^{(\gamma)} / \gamma$ as a function of $\lambda$ and $\mu/H$.
        The left panel shows the full parameter plane, while the right panel
        displays representative fixed-$\lambda$ slices.
    }
    \label{fig:equilateral_gamma}
\end{figure*}

Finally, using the Schwinger-Keldysh rules, the operator $\sigma^3$ gives
\begin{align}
B_\gamma(k_1,k_2,k_3)
&=
\frac{-2\gamma}{\sqrt{2\epsilon} m_{\rm Pl}}
\Im \!
\int_{-\infty(1-i0^+)}^{\tau_f} \!\!\!\!\!\!\!\!\!\!\!\!\!\!\!\!\!\!\!\!\!
\dd{\tau}\,
a^4(\tau) 
K_\sigma^+
\left(
k_1;\tau_f,\tau
\right)
\nn\\
&\quad\times
K_\sigma^+
\left(
k_2;\tau_f,\tau
\right)
K_\sigma^+
\left(
k_3;\tau_f,\tau
\right).
\label{eq-B-gamma-tau}
\end{align}
After complex conjugation and substitution of the dressed propagators, this becomes
\begin{align}
B_\gamma
&=
\frac{2\gamma H^2}{(2\epsilon)^2 m_{\rm Pl}^4}
\Im
\int_{-\infty(1+i0^+)}^{\tau_f} \!\!\!\!\!\!\!\!\!\!\!\!\!\!\!\!\!\!\!\!\!
\dd{\tau}\,
a^4(\tau)H \!\!\!\!\!\!\!\!\!
\sum_{w_1,w_2,w_3=\pm1}
\nn\\
&\quad\times
\mathcal F_U^{(w_1)*}(k_1,\tau_f)
\mathcal F_U^{(w_2)*}(k_2,\tau_f)
\mathcal F_U^{(w_3)*}(k_3,\tau_f)
\nn\\
&\quad\times
\mathcal S_U^{(w_1)}(k_1,\tau)
\mathcal S_U^{(w_2)}(k_2,\tau)
\mathcal S_U^{(w_3)}(k_3,\tau).
\label{eq-B-gamma-modes}
\end{align}
This expression is already symmetric under permutations of the three external momenta, so no additional terms are required. The corresponding reduced vertex integral is
\begin{align}
I_\gamma^{(\boldsymbol w)}(k_1,k_2,k_3)
&=
\int_{-\infty(1+i0^+)}^{\tau_f} \!\!\!\!\!\!\!\!\!\!\!\!\!\!\!\!\!\!\!\!\!
\dd{\tau}\,
\tau^2e^{-iK\tau}
\nn\\
& \!\!\!\!\!\!\!\!\!\!\!\!\!\!\!\!\!\!\!\!\! \times
h_U^{(w_1)}(-k_1\tau)
h_U^{(w_2)}(-k_2\tau)
h_U^{(w_3)}(-k_3\tau).
\end{align}
Inserting the integral representation~\eqref{h-int-2} for each of the three functions and using Eq.~\eqref{tau-int-Omega} gives
\begin{align}
I_\gamma^{(\boldsymbol w)}
&=
-i
\frac{
e^{\pi(w_1+w_2+w_3)\lambda/4}
}{
2\sqrt{2}\,
\Gamma(1+a_1)
\Gamma(1+a_2)
\Gamma(1+a_3)
}
\nn\\
&\quad\times
\int_0^\infty\dd{t_1}
\int_0^\infty\dd{t_2}
\int_0^\infty\dd{t_3}
\left[
\frac{2}{\Omega^3}
+
\frac{2i\tau_f}{\Omega^2}
-
\frac{\tau_f^2}{\Omega}
\right]
\nn\\
&\quad\times
e^{-i\tau_f\Omega}
D_\nu^{(w_1)}(-t_1)
D_\nu^{(w_2)}(-t_2)
D_\nu^{(w_3)}(-t_3)
\nn\\
&\quad\times
\left(
\frac{t_1}{1+t_1}
\right)^{a_1}
\left(
\frac{t_2}{1+t_2}
\right)^{a_2}
\left(
\frac{t_3}{1+t_3}
\right)^{a_3}.
\label{Igamma-result}
\end{align}
Figure~\ref{fig:equilateral_gamma} shows the normalized equilateral amplitude $f_{\rm NL}^{(\gamma)}/\gamma$ as a function of $\lambda$ and $\mu/H$. The left panel shows the full parameter plane, with the solid black curve denoting $f_{\rm NL}^{(\gamma)}/\gamma=0$ and the dotted diagonal marking $m_{\rm eff}^{2}=\mu^{2}-H^{2}\lambda^{2}=0$. The right panel shows fixed-$\lambda$ slices. The behavior near $(\lambda,\mu/H)=(0,0)$ is again nonuniform. In terms of $\Delta\equiv\frac32-\nu \simeq\frac{\mu^2}{3H^2}$, the leading behavior around the origin is
\begin{equation}
\frac{f_{\rm NL}^{(\gamma)}}{\gamma}
\simeq
\frac{\lambda^3}
{\left(\lambda^2+\Delta^2\right)^2}.
\label{eq:gamma-origin}
\end{equation}
Therefore, at fixed $\lambda>0$, this gives $ \lim_{\mu/H\to0} f_{\rm NL}^{(\gamma)}/\gamma = 1/\lambda$, whereas at fixed $\mu/H>0$ the amplitude vanishes as $\lambda^3$ when $\lambda\to0$. Along $\lambda=c\,\mu/H$, one instead finds $f_{\rm NL}^{(\gamma)}/\gamma\sim H/(c\mu)$, which diverges at the origin. The double limit therefore depends on the trajectory in parameter space.


\section{Squeezed limit of the bispectrum}
\label{sec:squeezed}

We now extract the squeezed limits of the bispectra derived in the preceding section. We specialize to the symmetric configuration in which $k_3$ is much smaller than the other two momenta:
\begin{align}
k_L&\equiv k_3,
\\
k_S&\equiv k_1=k_2.
\end{align}
The hierarchy is parametrized by
\be
r\equiv\frac{k_L}{k_S}\ll1.
\ee
Our strategy is to perform first the auxiliary integration associated with the long mode, which is $t_3$ when $k_3=k_L$. Inspection of Eqs.~\eqref{Ialpha-result}, \eqref{Ibeta-result}, and~\eqref{Igamma-result} shows that two types of long-mode integrals arise:
\begin{align}
\mathcal I_{\mathcal F}
\equiv
\int_0^\infty\dd{t}\,
&\left[
\frac{2}{\Omega^3}
+
\frac{2i\tau_f}{\Omega^2}
-
\frac{\tau_f^2}{\Omega}
\right]
e^{-i\tau_f\Omega}
\nn\\
&\quad\times
D_\nu^{(w)}(-t)
\frac{1}{t^2}
\left(
\frac{t}{1+t}
\right)^{1+a},
\label{I-int-g}
\end{align}
and
\begin{align}
\mathcal I_{\mathcal S}
\equiv
\int_0^\infty\dd{t}\,
&\left[
\frac{2}{\Omega^3}
+
\frac{2i\tau_f}{\Omega^2}
-
\frac{\tau_f^2}{\Omega}
\right]
e^{-i\tau_f\Omega}
\nn\\
&\quad\times
D_\nu^{(w)}(-t)
\left(
\frac{t}{1+t}
\right)^a,
\label{I-int-h}
\end{align}
where $a=iw\lambda/2$ and
\be
\Omega
\equiv
K+2k_S\left(t_i+t_j+rt\right).
\ee
Here $t_i$ and $t_j$ denote the two remaining auxiliary variables. The integral $\mathcal I_{\mathcal F}$ arises when the long leg carries $D_\tau\varphi$, whereas $\mathcal I_{\mathcal S}$ arises when it carries $\sigma$. As we shall see, $\mathcal I_{\mathcal S}$ gives the leading nonanalytic squeezed contribution, while $\mathcal I_{\mathcal F}$ is suppressed by two additional powers of $r$. 

\subsection{Long-mode integral}

The nonanalytic squeezed dependence originates from the region $t\sim r^{-1}$. Introducing $t=y/r$ makes the argument of $D_\nu^{(w)}$ large in this region, allowing us to use
\begin{align}
D_\nu^{(w)}
\left(
-\frac{y}{r}
\right)
&=
C_+^{(w)}
\left(
\frac{y}{r}
\right)^{\nu-\frac12}
\left[
1+\mathcal O\left(\frac{r}{y}\right)
\right]
\nn\\
&\quad+
C_-^{(w)}
\left(
\frac{y}{r}
\right)^{-\nu-\frac12}
\left[
1+\mathcal O\left(\frac{r}{y}\right)
\right],
\label{F3-exp}
\end{align}
where $C_\pm^{(w)}$ are defined in Eq.~\eqref{Cpm-def}. In the same region,
\begin{align}
\left(
\frac{y}{r+y}
\right)^A
=
1+\mathcal O\left(\frac{r}{y}\right).
\label{eq:approx-imaginary-power}
\end{align}
These expansions should be understood in the sense of the method of regions: they determine the nonanalytic dependence on $r$, whereas the complementary region $t=\mathcal O(1)$ contributes terms analytic in $r$. Thus, keeping the leading terms in the large-$t$ region gives
\begin{align}
\mathcal I_{\mathcal F}
\simeq
\int_0^\infty\dd{t}\,
&\left[
\frac{2}{\Omega^3}
+
\frac{2i\tau_f}{\Omega^2}
-
\frac{\tau_f^2}{\Omega}
\right]
e^{-i\tau_f\Omega}
\nn\\
&\quad\times
\left[
C_+^{(w)}t^{\nu-\frac52}
+
C_-^{(w)}t^{-\nu-\frac52}
\right],
\label{I-Z-power}
\end{align}
and
\begin{align}
\mathcal I_{\mathcal S}
\simeq
\int_0^\infty\dd{t}\,
&\left[
\frac{2}{\Omega^3}
+
\frac{2i\tau_f}{\Omega^2}
-
\frac{\tau_f^2}{\Omega}
\right]
e^{-i\tau_f\Omega}
\nn\\
&\quad\times
\left[
C_+^{(w)}t^{\nu-\frac12}
+
C_-^{(w)}t^{-\nu-\frac12}
\right].
\label{I-S-power}
\end{align}
Both expressions reduce to integrals of the form
\begin{align}
\mathcal I_\eta
&\equiv
\int_0^\infty\dd{t}\,
t^\eta
\left[
\frac{2}{\Omega^3}
+
\frac{2i\tau_f}{\Omega^2}
-
\frac{\tau_f^2}{\Omega}
\right]
e^{-i\tau_f\Omega},
\end{align}
where $\eta$ takes one of the values $-\frac12\pm\nu$ or $-\frac52\pm\nu$. Writing
\be
\Omega
=
k_S\left(J_r+2rt\right),
\qquad
J_r\equiv2(1+t_i+t_j)+r,
\label{Omega-hL-alpha}
\ee
and using the integral representation~\eqref{app-int-U}, one obtains
\begin{align}
\mathcal I_\eta
&=
k_S^{-3}
(2r)^{-\eta-1}
J_r^{\eta-2}
e^{-q}
\Gamma(\eta+1)
\nn\\
&\quad\times
\Big[
2U(\eta+1,\eta-1,q)
+
2q\,U(\eta+1,\eta,q)
\nn\\
&\hspace{29mm}
+
q^2U(\eta+1,\eta+1,q)
\Big],
\end{align}
where
\be
q\equiv i\tau_fk_SJ_r.
\ee
For $-\!1<\operatorname{Re}\eta<2$, the late-time limit can be taken directly. The resulting expression can then be analytically continued to the remaining values of $\eta$ required here, thereby isolating the nonanalytic contribution:
\begin{equation}
\mathcal I_\eta
=
k_S^{-3}
(2r)^{-\eta-1}
J_0^{\eta-2}
\Gamma(2-\eta)
\Gamma(\eta+1),
\label{eq:Ieta-approximated}
\end{equation}
where $J_0=2(1+t_i+t_j)$. It follows that
\begin{align}
\mathcal I_{\mathcal F}
&=
\frac{C_+^{(w)}}{k_S^3J_0^3}
\left(
\frac{2r}{J_0}
\right)^{\frac32-\nu}
\Gamma\left(\frac92-\nu\right)
\Gamma\left(\nu-\frac32\right)
\nn\\
&\quad+
\frac{C_-^{(w)}}{k_S^3J_0^3}
\left(
\frac{2r}{J_0}
\right)^{\frac32+\nu}
\Gamma\left(\frac92+\nu\right)
\Gamma\left(-\nu-\frac32\right),
\end{align}
whereas
\begin{align}
\mathcal I_{\mathcal S}
&=
\frac{C_+^{(w)}}{k_S^3J_0^3}
\left(
\frac{2r}{J_0}
\right)^{-\nu-\frac12}
\Gamma\left(\frac52-\nu\right)
\Gamma\left(\frac12+\nu\right)
\nn\\
&\quad+
\frac{C_-^{(w)}}{k_S^3J_0^3}
\left(
\frac{2r}{J_0}
\right)^{\nu-\frac12}
\Gamma\left(\frac52+\nu\right)
\Gamma\left(\frac12-\nu\right).
\end{align}
Thus, the leading nonanalytic contribution comes from $\mathcal I_{\mathcal S}$ and scales as $r^{-\nu-1/2}$ at the level of the reduced vertex integral. After including the kinematic factor in the definition of the shape, this becomes the characteristic scaling $r^{1/2-\nu}$. By contrast, the contribution from $\mathcal I_{\mathcal F}$ begins at $r^{5/2-\nu}$ in the shape and is subleading.

\subsection{$\alpha$ interaction}

We begin with the interaction proportional to $\alpha$. In the configuration $k_3=k_L\ll k_1=k_2=k_S$, the leading nonanalytic contribution arises when the entropy leg carries the long momentum. Equation~\eqref{Ialpha-result} then becomes
\begin{align}
I_\alpha^{(\boldsymbol w)}
&=
-i
\frac{
e^{\pi(w_1+w_2+w_3)\lambda/4}
}{
2\sqrt{2}\,
\Gamma(a_1)\Gamma(a_2)\Gamma(1+a_3)
}
\nn\\
&\quad\times
\int_0^\infty\dd{t_1}
\int_0^\infty\dd{t_2}\,
D_\nu^{(w_1)}(-t_1)
D_\nu^{(w_2)}(-t_2)
\nn\\
&\quad\times
\frac{1}{t_1^2}
\left(
\frac{t_1}{1+t_1}
\right)^{1+a_1}
\frac{1}{t_2^2}
\left(
\frac{t_2}{1+t_2}
\right)^{1+a_2}
\mathcal I_{\mathcal S}^{(w_3)}.
\label{Ialpha-result-2}
\end{align}
For $J_0=2(1+t_1+t_2)$, the long-mode integral is
\begin{align}
\mathcal I_{\mathcal S}^{(w_3)}
&=
\frac{1}{8k_S^3}
\Bigg[
C_+^{(w_3)}
\Gamma\left(\frac52-\nu\right)
\Gamma\left(\frac12+\nu\right)
\nn\\
&\hspace{17mm}\times
(1+t_1+t_2)^{\nu-\frac52}
r^{-\nu-\frac12}
\nn\\
&\quad+
C_-^{(w_3)}
\Gamma\left(\frac52+\nu\right)
\Gamma\left(\frac12-\nu\right)
\nn\\
&\hspace{17mm}\times
(1+t_1+t_2)^{-\nu-\frac52}
r^{\nu-\frac12}
\Bigg].
\label{eq:Is-alpha-squeezed}
\end{align}
The reduced vertex integral can therefore be organized as
\begin{align}
I_\alpha^{(\boldsymbol w)}(k_S,k_S,rk_S)
&=
-\frac{
ie^{\pi(w_1+w_2+w_3)\lambda/4}
}{
16\sqrt{2}\,k_S^3\,
\Gamma(a_1)\Gamma(a_2)\Gamma(1+a_3)
}
\nn\\
&\quad\times
\left[
\mathcal I_{\alpha,+}^{(\boldsymbol w)}
r^{-\nu-\frac12}
+
\mathcal I_{\alpha,-}^{(\boldsymbol w)}
r^{\nu-\frac12}
\right]
+\cdots,
\label{eq:Ialpha-squeezed}
\end{align}
where
\begin{align}
\mathcal I_{\alpha,+}^{(\boldsymbol w)}
&\equiv
C_+^{(w_3)}
\Gamma\left(\frac52-\nu\right)
\Gamma\left(\frac12+\nu\right)
\mathcal J_{\alpha,+}^{(w_1,w_2)},
\\
\mathcal I_{\alpha,-}^{(\boldsymbol w)}
&\equiv
C_-^{(w_3)}
\Gamma\left(\frac52+\nu\right)
\Gamma\left(\frac12-\nu\right)
\mathcal J_{\alpha,-}^{(w_1,w_2)},
\end{align}
and
\begin{align}
\mathcal J_{\alpha,\pm}^{(w_1,w_2)}
&\equiv
\int_0^\infty\dd{t_1}
\int_0^\infty\dd{t_2}\,
D_\nu^{(w_1)}(-t_1)
D_\nu^{(w_2)}(-t_2)
\nn\\
&\quad\times
\frac{1}{t_1^2}
\left(
\frac{t_1}{1+t_1}
\right)^{1+a_1}
\frac{1}{t_2^2}
\left(
\frac{t_2}{1+t_2}
\right)^{1+a_2}
\nn\\
&\quad\times
(1+t_1+t_2)^{-\frac52\pm\nu}.
\label{eq:Jalpha-pm}
\end{align}
Using the late-time limit
\begin{equation}
\mathcal F_U^{(w)}(k,0)
=
\frac{H}{\sqrt{2k^3}}\,
\mathcal F_0^{(w)},
\end{equation}
with $\mathcal F_0^{(w)}$ given in Eq.~\eqref{g0-integral-short}, the nonanalytic part of the dimensionless shape becomes
\begin{align}
\mathcal S_\alpha(r;\nu,\lambda)
&=
-\frac{2\alpha}{(2\epsilon)^2}
\frac{(H/m_{\rm Pl})^4}{
2^{10}\sqrt{2}\,\pi^4\Delta_\zeta^2
}
\nn\\
&\quad\times
\left[
c_{\alpha,+}(\nu,\lambda)
r^{\frac12-\nu}
+
c_{\alpha,-}(\nu,\lambda)
r^{\frac12+\nu}
\right],
\label{eq:Salpha-squeezed}
\end{align}
where
\be
c_{\alpha,\pm}(\nu,\lambda)
\equiv
\Im \!\!\!\!\!\!\!\!\!\!
\sum_{w_1,w_2,w_3=\pm1} \!\!\!\!\!\!\!\!\!\!
\mathcal F_0^{(w_1)*}
\mathcal F_0^{(w_2)*}
\mathcal F_0^{(w_3)*}
\mathcal N_\alpha^{(\boldsymbol w)}
\mathcal I_{\alpha,\pm}^{(\boldsymbol w)},
\label{eq:calpha-pm}
\ee
with
\begin{equation}
\mathcal N_\alpha^{(\boldsymbol w)}
\equiv
-\frac{
iw_3e^{\pi(w_1+w_2+w_3)\lambda/4}
}{
\Gamma(a_1)\Gamma(a_2)\Gamma(1+a_3)
}.
\label{eq:Nalpha}
\end{equation}
For $0<\nu<3/2$, the term proportional to $r^{1/2-\nu}$ gives the leading nonanalytic squeezed behavior.

\subsection{$\beta$ interaction}

For the interaction proportional to $\beta$, the leading signal arises whenever the long momentum is carried by either of the two entropy legs. In the symmetric squeezed configuration these two contributions are equal, producing an overall factor of two. Choosing $k_3=k_L$ for the intermediate calculation, Eq.~\eqref{Ibeta-result} becomes
\begin{align}
I_\beta^{(\boldsymbol w)}
&=
-i
\frac{
e^{\pi(w_1+w_2+w_3)\lambda/4}
}{
2\sqrt{2}\,
\Gamma(a_1)\Gamma(1+a_2)\Gamma(1+a_3)
}
\nn\\
&\quad\times
\int_0^\infty\dd{t_1}
\int_0^\infty\dd{t_2}\,
D_\nu^{(w_1)}(-t_1)
D_\nu^{(w_2)}(-t_2)
\nn\\
&\quad\times
\frac{1}{t_1^2}
\left(
\frac{t_1}{1+t_1}
\right)^{1+a_1}
\left(
\frac{t_2}{1+t_2}
\right)^{a_2}
\mathcal I_{\mathcal S}^{(w_3)}.
\end{align}
It follows that
\begin{align}
I_\beta^{(\boldsymbol w)}(k_S,k_S,rk_S)
&=
-\frac{
ie^{\pi(w_1+w_2+w_3)\lambda/4}
}{
16\sqrt{2}\,k_S^3\,
\Gamma(a_1)\Gamma(1+a_2)\Gamma(1+a_3)
}
\nn\\
&\quad\times
\left[
\mathcal I_{\beta,+}^{(\boldsymbol w)}
r^{-\nu-\frac12}
+
\mathcal I_{\beta,-}^{(\boldsymbol w)}
r^{\nu-\frac12}
\right]
+\cdots,
\label{eq:Ibeta-squeezed}
\end{align}
where
\begin{align}
\mathcal I_{\beta,+}^{(\boldsymbol w)}
&\equiv
C_+^{(w_3)}
\Gamma\left(\frac52-\nu\right)
\Gamma\left(\frac12+\nu\right)
\mathcal J_{\beta,+}^{(w_1,w_2)},
\\
\mathcal I_{\beta,-}^{(\boldsymbol w)}
&\equiv
C_-^{(w_3)}
\Gamma\left(\frac52+\nu\right)
\Gamma\left(\frac12-\nu\right)
\mathcal J_{\beta,-}^{(w_1,w_2)},
\end{align}
and
\begin{align}
\mathcal J_{\beta,\pm}^{(w_1,w_2)}
&\equiv
\int_0^\infty\dd{t_1}
\int_0^\infty\dd{t_2}\,
D_\nu^{(w_1)}(-t_1)
D_\nu^{(w_2)}(-t_2)
\nn\\
&\quad\times
\frac{1}{t_1^2}
\left(
\frac{t_1}{1+t_1}
\right)^{1+a_1}
\left(
\frac{t_2}{1+t_2}
\right)^{a_2}
\nn\\
&\quad\times
(1+t_1+t_2)^{-\frac52\pm\nu}.
\label{eq:Jbeta-pm}
\end{align}
The resulting nonanalytic shape is
\begin{align}
\mathcal S_\beta(r;\nu,\lambda)
&=
-\frac{2\beta}{(2\epsilon)^2}
\frac{(H/m_{\rm Pl})^4}{
2^8\sqrt{2}\,\pi^4\Delta_\zeta^2
}
\nn\\
&\quad\times
\left[
c_{\beta,+}(\nu,\lambda)
r^{\frac12-\nu}
+
c_{\beta,-}(\nu,\lambda)
r^{\frac12+\nu}
\right],
\label{eq:Sbeta-squeezed}
\end{align}
where
\be
c_{\beta,\pm}(\nu,\lambda)
\equiv
\Im \!\!\!\!\!\!\!\!\!\!
\sum_{w_1,w_2,w_3=\pm1} \!\!\!\!\!\!\!\!\!\!
\mathcal F_0^{(w_1)*}
\mathcal F_0^{(w_2)*}
\mathcal F_0^{(w_3)*}
\mathcal N_\beta^{(\boldsymbol w)}
\mathcal I_{\beta,\pm}^{(\boldsymbol w)},
\label{eq:cbeta-pm}
\ee
with
\begin{equation}
\mathcal N_\beta^{(\boldsymbol w)}
\equiv
\frac{
w_2w_3e^{\pi(w_1+w_2+w_3)\lambda/4}
}{
\Gamma(a_1)\Gamma(1+a_2)\Gamma(1+a_3)
}.
\label{eq:Nbeta}
\end{equation}
Again, the leading nonanalytic term for $0<\nu<3/2$ scales as $r^{1/2-\nu}$.

\subsection{$\gamma$ interaction}

Finally, for the interaction proportional to $\gamma$, all three bulk legs are entropy modes. Choosing $k_3=k_L$, the reduced integral~\eqref{Igamma-result} becomes
\begin{align}
I_\gamma^{(\boldsymbol w)}
&=
-\frac{
ie^{\pi(w_1+w_2+w_3)\lambda/4}
}{
16\sqrt{2}\,k_S^3\,
\Gamma(1+a_1)
\Gamma(1+a_2)
\Gamma(1+a_3)
}
\nn\\
&\quad\times
\left[
\mathcal I_{\gamma,+}^{(\boldsymbol w)}
r^{-\nu-\frac12}
+
\mathcal I_{\gamma,-}^{(\boldsymbol w)}
r^{\nu-\frac12}
\right]
+\cdots,
\label{eq:Igamma-squeezed}
\end{align}
where
\begin{align}
\mathcal I_{\gamma,+}^{(\boldsymbol w)}
&\equiv
C_+^{(w_3)}
\Gamma\left(\frac52-\nu\right)
\Gamma\left(\frac12+\nu\right)
\mathcal J_{\gamma,+}^{(w_1,w_2)},
\\
\mathcal I_{\gamma,-}^{(\boldsymbol w)}
&\equiv
C_-^{(w_3)}
\Gamma\left(\frac52+\nu\right)
\Gamma\left(\frac12-\nu\right)
\mathcal J_{\gamma,-}^{(w_1,w_2)},
\end{align}
and
\begin{align}
\mathcal J_{\gamma,\pm}^{(w_1,w_2)}
&\equiv
\int_0^\infty\dd{t_1}
\int_0^\infty\dd{t_2}\,
D_\nu^{(w_1)}(-t_1)
D_\nu^{(w_2)}(-t_2)
\nn\\
&\quad\times
\left(
\frac{t_1}{1+t_1}
\right)^{a_1}
\left(
\frac{t_2}{1+t_2}
\right)^{a_2}
(1+t_1+t_2)^{-\frac52\pm\nu}.
\label{eq:Jgamma-pm}
\end{align}
The corresponding nonanalytic shape is
\begin{align}
\mathcal S_\gamma(r;\nu,\lambda)
&=
\frac{2\gamma}{(2\epsilon)^2}
\frac{(H/m_{\rm Pl})^4}{
2^8\sqrt{2}\,\pi^4\Delta_\zeta^2
}
\nn\\
&\quad\times
\left[
c_{\gamma,+}(\nu,\lambda)
r^{\frac12-\nu}
+
c_{\gamma,-}(\nu,\lambda)
r^{\frac12+\nu}
\right],
\label{eq:Sgamma-squeezed}
\end{align}
where
\be
c_{\gamma,\pm}(\nu,\lambda)
\equiv
\Im \!\!\!\!\!\!\!\!\!\!
\sum_{w_1,w_2,w_3=\pm1} \!\!\!\!\!\!\!\!\!\!
\mathcal F_0^{(w_1)*}
\mathcal F_0^{(w_2)*}
\mathcal F_0^{(w_3)*}
\mathcal N_\gamma^{(\boldsymbol w)}
\mathcal I_{\gamma,\pm}^{(\boldsymbol w)},
\label{eq:cgamma-pm}
\ee
with
\begin{equation}
\mathcal N_\gamma^{(\boldsymbol w)}
\equiv
-\frac{
iw_1w_2w_3
e^{\pi(w_1+w_2+w_3)\lambda/4}
}{
\Gamma(1+a_1)
\Gamma(1+a_2)
\Gamma(1+a_3)
}.
\label{eq:Ngamma}
\end{equation}

The three interactions therefore produce the same pair of nonanalytic squeezed exponents, $r^{1/2-\nu}$ and $r^{1/2+\nu}$, while differing in their amplitudes and dependence on $\lambda$. For a heavy entropy field, analytic continuation to $\nu=i\rho$ combines the two branches into logarithmic oscillations proportional to $r^{1/2}\sin[\rho\ln(1/r)+\phi]$, with an interaction-dependent amplitude and phase.

\subsection{Strong-mixing asymptotics}
\label{sec:strong-mixing-asymptotics}

We conclude this section by examining the squeezed-limit coefficients in the strong-mixing regime. We take $\lambda\to+\infty$ at fixed, generic $\nu$, with $2\nu\notin\mathbb Z$. The case of large negative $\lambda$ follows by interchanging the two branches, $w=+1\leftrightarrow-1$. The asymptotic analysis requires a uniform treatment of the auxiliary integrals. Although $D_\nu^{(w)}(-t)\to1$ at fixed $t$ as $\lambda\to\infty$, the relevant integration region scales as $t\sim\lambda$. Setting $t=\lambda y$, we define
\be
D_\nu^{(w)}(-\lambda y)
=
\mathcal D_{\nu,\infty}^{(w)}(y)
+
\mathcal O(\lambda^{-1}).
\label{eq:D-large-lambda-uniform}
\ee
The hypergeometric equation satisfied by $D_\nu^{(w)}$ then reduces to
\be
\left[
y^2\dv[2]{y}
+
(iw+2y)\dv{y}
+
\frac14-\nu^2
\right]
\mathcal D_{\nu,\infty}^{(w)}(y)
=
0.
\label{eq:D-large-lambda-ode}
\ee
Matching to $D_\nu^{(w)}(0)=1$ selects the solutions
\begin{align}
\mathcal D_{\nu,\infty}^{(+)}(y)
&=
\frac{\sqrt{\pi}}{2}
e^{-\frac{i\pi}{4}(2\nu+1)}
y^{-1/2}e^{\frac{i}{2y}}
H_\nu^{(2)}
\left(\frac{1}{2y}\right), 
\\
\mathcal D_{\nu,\infty}^{(-)}(y)
&=
\frac{\sqrt{\pi}}{2}
e^{\frac{i\pi}{4}(2\nu+1)}
y^{-1/2}e^{-\frac{i}{2y}}
H_\nu^{(1)}
\left(\frac{1}{2y}\right).
\label{eq:D-large-lambda-hankel}
\end{align}
At the same time $
\left(t / (1+t)
\right)^{\frac{iw\lambda}{2}}
\longrightarrow
e^{-\frac{iw}{2y}}$. The two types of kernels appearing in the hard-mode integrals then behave as
\begin{align}
\dd{t}
D_\nu^{(w)}(-t)
\frac{1}{t^2}
\left(
\frac{t}{1+t}
\right)^{1+\frac{iw\lambda}{2}}  =& \,
\lambda^{-1}\dd{y}\,
y^{-2} e^{-\frac{iw}{2y}} \nn \\
& \!\!\!\!\!\!\!\!\!\!\!\!\!\!\!\!\!\!\!\!\!\!\!\!\!\!\!\! \times
\mathcal D_{\nu,\infty}^{(w)}(y)
\left[
1+\mathcal O(\lambda^{-1})
\right], \quad
\\
 \dd{t}
D_\nu^{(w)}(-t)
\left(
\frac{t}{1+t}
\right)^{\frac{iw\lambda}{2}} = & \,
\lambda\,\dd{y}\,
e^{-\frac{iw}{2y}}  \nn \\
& \!\!\!\!\!\!\!\!\!\!\!\!\!\!\!\!\!\!\!\!\!\!\!\!\!\!\!\!  \times
\mathcal D_{\nu,\infty}^{(w)}(y)
\left[
1+\mathcal O(\lambda^{-1})
\right].
\label{eq:kernel-large-lambda}
\end{align}
Thus, every $g$-type leg contributes a factor $\lambda^{-1}$, whereas every $h$-type leg contributes a factor $\lambda$. Since $1+t_1+t_2\sim\lambda(y_1+y_2)$, the integrals defined in Eqs.~\eqref{eq:Jalpha-pm}, \eqref{eq:Jbeta-pm}, and~\eqref{eq:Jgamma-pm} satisfy
\begin{align}
\mathcal J_{\alpha,\pm}^{(w_1,w_2)}
&=
\lambda^{-\frac92\pm\nu}
\left[
\widehat{\mathcal J}_{\alpha,\pm}^{(w_1,w_2)}
+
\mathcal O(\lambda^{-1})
\right],
\\
\mathcal J_{\beta,\pm}^{(w_1,w_2)}
&=
\lambda^{-\frac52\pm\nu}
\left[
\widehat{\mathcal J}_{\beta,\pm}^{(w_1,w_2)}
+
\mathcal O(\lambda^{-1})
\right],
\\
\mathcal J_{\gamma,\pm}^{(w_1,w_2)}
&=
\lambda^{-\frac12\pm\nu}
\left[
\widehat{\mathcal J}_{\gamma,\pm}^{(w_1,w_2)}
+
\mathcal O(\lambda^{-1})
\right],
\label{eq:J-large-lambda}
\end{align}
where the hatted quantities are finite and independent of $\lambda$.

The coefficients in the large-$t$ expansion of the dressing function behave as
\be
C_\pm^{(w)}
=
\frac{\Gamma(\pm2\nu)}
{\Gamma(\frac12\pm\nu)}
(iw\lambda)^{\frac12\mp\nu}
\left[
1+\mathcal O(\lambda^{-1})
\right].
\label{eq:Cpm-large-lambda}
\ee
The explicit powers of $\nu$ consequently cancel between $C_\pm^{(w)}$ and $\mathcal J_{I,\pm}$. At the level of power counting, this gives
\begin{align}
\mathcal I_{\alpha,\pm}^{(\boldsymbol w)}
&=
\mathcal O(\lambda^{-4}), \label{eq:I-large-lambda-power-counting-1} \\
\mathcal I_{\beta,\pm}^{(\boldsymbol w)}
&=
\mathcal O(\lambda^{-2}), \label{eq:I-large-lambda-power-counting-2} \\
\mathcal I_{\gamma,\pm}^{(\boldsymbol w)}
&=
\mathcal O(1).
\label{eq:I-large-lambda-power-counting-3}
\end{align}
The remaining factors select the branch $\boldsymbol w=(+++)$ exponentially. Indeed
\begin{align}
\mathcal F_0^{(+)}
&=
\mathcal O\left(
e^{\pi\lambda/2}\lambda^{-1/2}
\right), 
\\
\mathcal F_0^{(-)}
&=
\mathcal O\left(
\lambda^{-1/2}
\right),
\end{align}
while
\begin{align}
\mathcal N_\alpha^{(+++)}
&=
\mathcal O\left(
e^{3\pi\lambda/2}\lambda^{1/2}
\right),
\nn\\
\mathcal N_\beta^{(+++)}
&=
\mathcal O\left(
e^{3\pi\lambda/2}\lambda^{-1/2}
\right),
\nn\\
\mathcal N_\gamma^{(+++)}
&=
\mathcal O\left(
e^{3\pi\lambda/2}\lambda^{-3/2}
\right).
\label{eq:N-large-lambda}
\end{align}
A direct combination of these estimates would appear to give $c_{\alpha,\pm}\propto e^{3\pi\lambda}\lambda^{-5}$, $c_{\beta,\pm}\propto e^{3\pi\lambda}\lambda^{-4}$, and $c_{\gamma,\pm}\propto e^{3\pi\lambda}\lambda^{-3}$. These leading terms, however, cancel after taking the imaginary part in Eqs.~\eqref{eq:calpha-pm}, \eqref{eq:cbeta-pm}, and~\eqref{eq:cgamma-pm}. To see this cancellation, define
\be
p_\pm
\equiv
\frac52\mp\nu.
\ee
After changing variables to $x_i=1/(2y_i)$, the contours of the leading integrals may be rotated to $x_i=-iu_i$. The Hankel functions are thereby converted into modified Bessel functions, leaving real Euclidean integrals multiplied by the phases
\begin{align}
\arg\widehat{\mathcal J}_{\alpha,\pm}^{(+,+)}
&=
-\frac{\pi}{2}(p_\pm+2),
\nn\\
\arg\widehat{\mathcal J}_{\beta,\pm}^{(+,+)}
&=
-\frac{\pi}{2}p_\pm,
\nn\\
\arg\widehat{\mathcal J}_{\gamma,\pm}^{(+,+)}
&=
-\frac{\pi}{2}(p_\pm-2),
\label{eq:J-large-lambda-phases}
\end{align}
up to integer multiples of $\pi$. On the other hand
\be
\arg C_\pm^{(+)}
=
\frac{\pi}{4}
\mp
\frac{\pi\nu}{2}
\qquad
\operatorname{mod}\pi.
\ee
It follows that
\begin{align}
C_\pm^{(+)}
\widehat{\mathcal J}_{\alpha,\pm}^{(+,+)}
&\in\mathbb R, \\
C_\pm^{(+)}
\widehat{\mathcal J}_{\beta,\pm}^{(+,+)}
&\in\mathbb R,\\
C_\pm^{(+)}
\widehat{\mathcal J}_{\gamma,\pm}^{(+,+)}
&\in\mathbb R.
\end{align}
The leading products of the external modes with $\mathcal N_I^{(+++)}$ are likewise real. Hence
\be
\Im
\left[
\left(
\mathcal F_0^{(+)*}
\right)^3
\mathcal N_I^{(+++)}
\mathcal I_{I,\pm}^{(+++)}
\right]_{\rm leading}
=
0,
\label{eq:leading-large-lambda-cancellation}
\ee
for $I=\alpha,\beta,\gamma$. The leading power-counting contributions therefore vanish identically. Assuming a regular uniform expansion in inverse powers of $\lambda$, the first potentially nonzero terms obey the bounds
\begin{align}
c_{\alpha,\pm}(\nu,\lambda)
&=
\mathcal O\left(
\frac{e^{3\pi\lambda}}{\lambda^6}
\right),
\\
c_{\beta,\pm}(\nu,\lambda)
&=
\mathcal O\left(
\frac{e^{3\pi\lambda}}{\lambda^5}
\right), 
\\
c_{\gamma,\pm}(\nu,\lambda)
&=
\mathcal O\left(
\frac{e^{3\pi\lambda}}{\lambda^4}
\right).
\label{eq:c-large-lambda-bound}
\end{align}
Determining whether these bounds are saturated requires carrying the uniform expansion to the next order. Further cancellations at that order are not excluded by the leading analysis.

For comparison, the exact power spectrum~\eqref{power-zeta} behaves as
\begin{align}
\Delta_\zeta
= &
\frac{H^2}{16\pi^4\epsilon m_{\rm Pl}^2}
\left|
\Gamma\left(\frac34-\frac{\nu}{2}\right)
\Gamma\left(\frac34+\frac{\nu}{2}\right)
\right|^2 \nn 
\\ & \times
\frac{e^{\pi\lambda}}{\lambda}
\left[
1+\mathcal O(\lambda^{-1})
\right].
\label{eq:PS-large-lambda}
\end{align}
After accounting for the factor $\Delta_\zeta^{-2}$ in the definition of the dimensionless shape, Eq.~\eqref{eq:c-large-lambda-bound} implies
\begin{align}
\mathcal S_\alpha
&=
\mathcal O\left(
\frac{e^{\pi\lambda}}{\lambda^4}
\right), 
\\
\mathcal S_\beta
&=
\mathcal O\left(
\frac{e^{\pi\lambda}}{\lambda^3}
\right),
\\
\mathcal S_\gamma
&=
\mathcal O\left(
\frac{e^{\pi\lambda}}{\lambda^2}
\right),
\label{eq:S-large-lambda-bound}
\end{align}
at fixed $\nu$, background quantities, and covariant cubic couplings. These expressions should be understood as upper bounds rather than established asymptotic equalities. At degenerate values $2\nu\in\mathbb Z$, additional logarithmic factors may arise. Of course, taking $\lambda \gg 1$ while keeping $\nu$ (or, equivalently, $\mu$) fixed is only one possible way to probe the asymptotic behavior of the shape amplitudes. Other, equally interesting scaling regimes are expected to emerge when combined directions in parameter space are explored.


\section{Conclusions}

In this work, we developed an exact analytic framework for the study of primordial non-Gaussianity in a two-field inflationary system with constant curvature--isocurvature mixing. By incorporating the quadratic mixing directly into the mode functions, we obtained solutions valid nonperturbatively in $\lambda$ and for arbitrary values of the entropy mass $\mu$. These solutions, first presented in Ref.~\cite{Huenupi:2026abj}, allowed us to recover the exact power spectrum, construct dressed Schwinger--Keldysh propagators, and derive integral representations for the tree-level bispectra generated by the three cubic interactions in Eq.~\eqref{S3-covariant}.

The resulting squeezed limits exhibit the characteristic nonanalytic powers $r^{1/2\pm\nu}$, with exponents determined entirely by the entropy mass and coefficients that retain the full, nonperturbative dependence on the mixing strength. Analytic continuation to imaginary $\nu$ gives the logarithmic oscillations associated with heavy degrees of freedom. In the weak-mixing limit, these expressions reduce to the familiar quasi-single-field and cosmological-collider results. At strong mixing, they show instead that the amplitudes of both the power spectrum and the bispectrum can be substantially enhanced. They also make explicit that the frequency of the cosmological-collider signal is controlled by the entropy mass $\mu$, rather than by the effective mass $m_{\rm eff}$ obtained after separating the quadratic mixing from the free action.

Beyond these phenomenological results, the analytic construction reveals a structural property that may be of broader significance. The solutions at general $\nu$ need not be found independently for each value of the mass. Instead, the conformally coupled sector, $\nu=1/2$, supplies a set of seed solutions from which the complete family at arbitrary $\nu$ is generated by the mass-dressing operator $\widehat{\mathcal D}_{\nu}^{(w)}$. For half-integer values of $\nu$, this construction reduces to the repeated action of the raising operators of the $\mathfrak{sl}(2)$ ladder algebra, while the hypergeometric form of $\widehat{\mathcal D}_{\nu}^{(w)}$ provides its continuation to general mass. The conformally coupled theory is therefore not merely an isolated solvable case: it acts as a generating point for the entire spectrum of massive solutions.

This structure suggests a possible connection with the cosmological-bootstrap program, where correlators of general fields are likewise constructed from simpler conformally coupled seeds through weight-shifting and related differential operators~\cite{Arkani-Hamed:2018kmz, Baumann:2019oyu, Pajer:2020wxk, Pimentel:2022fsc, XianyuZhang:2022, Wang:2022eop, Jazayeri:2022kjy, Aoki:2024uyi, Liu:2024xyi}. The parallel is not yet an identification. The operator introduced here acts on bulk mode functions in a system with curvature--isocurvature mixing resummed to all orders, whereas bootstrap operators are usually formulated directly at the level of boundary correlators and their conformal weights. Nevertheless, it is natural to ask whether the mass-dressing operation admits a boundary counterpart and whether it can be used to derive recursion relations or differential constraints for the corresponding cosmological correlators. Such a formulation could provide a route toward bootstrapping inflationary observables in regimes of strong mixing, where conventional perturbative treatments of the transfer are no longer available.

It would also be interesting to extend the present construction to time-dependent mixing, higher-point functions, and loop corrections. More fundamentally, however, the results obtained here indicate that the conformally coupled sector contains considerably more information than its special mass value might suggest: once supplemented by the appropriate dressing operator, it encodes the dynamics of a much broader family of strongly mixed inflationary systems. \\


\noindent {\bf Note added.---}
While this manuscript was nearing completion, we became aware of forthcoming works by Lucas Pinol~\cite{Pinol:2026xnl} and Xiangwei Wang, Yi Wang and Yunke Zhao~\cite{Wang:2026lff}, inspired by the exact solutions presented in Ref.~\cite{Huenupi:2026abj}, which independently derive the squeezed limit of the bispectrum. 

\begin{acknowledgments}

\vspace{3pt}
We wish to thank Ana Ach\'ucarro, Perseas Christodoulidis, Gabriel Mar\'in Mac\^edo, Hayden Lee, Anish Pandya, Nicol\'as Parra, S\'ebastien Renaux-Petel, Lucas Pinol, Diederik Roest, Zhong-Zhi Xianyu and Crist\'obal Zenteno for useful discussions and comments. JH acknowledges the hospitality of the Munich Institute for Astro-, Particle and BioPhysics (MIAPbP) which is funded by the Deutsche Forschungsgemeinschaft (DFG, German Research
Foundation) under Germany’s Excellence Strategy – EXC-2094 – 390783311. CM acknowledges support from FONDECYT 1231250 and Centro de Modelamiento Matem\'atico (CMM) BASAL fund FB210005, and support from PRISMALab Group at CMM. GAP acknowledges support from the Fondecyt Regular projects 1210876 and 1251511 (ANID). SS is supported by FAPEI [grant number FP2688125].

\end{acknowledgments}

\begin{appendix}

\section{Tricomi and Kummer functions}
\label{app:Tri-Kum}

In this appendix, we collect the properties of the Kummer and Tricomi functions used throughout the main text. The confluent hypergeometric equation is
\be
x\frac{\dd^2F}{\dd x^2}
+
(b-x)\frac{\dd F}{\dd x}
-
aF
=
0.
\ee
Two standard solutions are Kummer's function $M(a,b,x)$ and Tricomi's function $U(a,b,x)$. The former is defined by
\be
M(a,b,x)
\equiv
{}_1F_1(a;b;x)
=
\sum_{n=0}^{\infty}
\frac{(a)_n}{(b)_n}
\frac{x^n}{n!},
\ee
where $(a)_n=\Gamma(a+n)/\Gamma(a)$ is the Pochhammer symbol. This solution is regular at $x=0$ provided that $b\neq0,-1,-2,\ldots$.

Tricomi's function is characterized by the large-$x$ behavior
\be
U(a,b,x)
\sim
x^{-a},
\qquad
|x|\to\infty,
\ee
on the principal branch and away from the branch cut, conventionally chosen along the negative real axis. For $b\notin\mathbb Z$, it can be expressed as the linear combination
\begin{align}
U(a,b,x)
&=
\frac{\Gamma(1-b)}{\Gamma(a-b+1)}
M(a,b,x)
\nn\\
&\quad+
\frac{\Gamma(b-1)}{\Gamma(a)}
x^{1-b}
M(a-b+1,2-b,x).
\end{align}
For integer $b$, this representation is understood by taking the appropriate limit. The two Frobenius branches then degenerate, giving rise to logarithmic terms. For example,
\begin{align}
U(a,1,x)
&=
-\frac{1}{\Gamma(a)}
\sum_{n=0}^{\infty}
\frac{(a)_n}{(n!)^2}
x^n
\nn\\
&\quad\times
\left[
\ln x
+
\psi(a+n)
-
2\psi(n+1)
\right],
\end{align}
and hence
\be
U(a,1,x)
=
-\frac{1}{\Gamma(a)}
\left[
\ln x
+
\psi(a)
+
2\gamma_{\rm E}
\right]
+
\mathcal O(x\ln x)
\ee
as $x\to0$. Similarly, the limit $b\to2$ gives
\begin{align}
U(a,2,x)
&=
\frac{1}{\Gamma(a)}\frac{1}{x}
+
\frac{1}{\Gamma(a-1)}
\sum_{n=0}^{\infty}
\frac{(a)_n}{(2)_n\,n!}
x^n
\nn\\
&\quad\times
\left[
\ln x
+
\psi(a+n)
-
\psi(n+1)
-
\psi(n+2)
\right].
\end{align}
Its small-$x$ behavior is therefore
\begin{align}
U(a,2,x)
&=
\frac{1}{\Gamma(a)}\frac{1}{x}
+
\frac{a-1}{\Gamma(a)}
\left[
\ln x
+
\psi(a)
+
2\gamma_{\rm E}
-
1
\right]
\nn\\
&\quad+
\mathcal O(x\ln x).
\end{align}

For suitable values of their parameters, both functions admit useful integral representations. When $\operatorname{Re}(b)>\operatorname{Re}(a)>0$, Kummer's function can be written as
\be
M(a,b,x)
=
\frac{\Gamma(b)}
{\Gamma(a)\Gamma(b-a)}
\int_0^1\dd{t}\,
e^{xt}
t^{a-1}
(1-t)^{b-a-1}.
\ee
For $\operatorname{Re}(a)>0$ and $\operatorname{Re}(x)>0$, Tricomi's function admits the representation
\be
U(a,b,x)
=
\frac{1}{\Gamma(a)}
\int_0^\infty\dd{t}\,
e^{-xt}
t^{a-1}
(1+t)^{b-a-1}.
\label{app-int-U}
\ee
Outside these convergence domains, the corresponding functions are obtained by analytic continuation once the contour and branch prescriptions have been specified.

The integral representations also make manifest the transformation identities
\begin{align}
M(a,b,x)
&=
e^xM(b-a,b,-x),
\\
U(a,b,x)
&=
x^{1-b}
U(1+a-b,2-b,x).
\end{align}
A special case of the second relation, used repeatedly in the main text, is
\be
U(a+1,2,x)
=
\frac{1}{x}U(a,0,x).
\label{U_id_b=2}
\ee

Differentiation shifts both parameters:
\begin{align}
\frac{\dd}{\dd x}M(a,b,x)
&=
\frac{a}{b}
M(a+1,b+1,x),
\\
\frac{\dd}{\dd x}U(a,b,x)
&=
-a\,U(a+1,b+1,x).
\end{align}
More generally,
\begin{align}
\frac{\dd^n}{\dd x^n}M(a,b,x)
&=
\frac{(a)_n}{(b)_n}
M(a+n,b+n,x),
\\
\frac{\dd^n}{\dd x^n}U(a,b,x)
&=
(-1)^n(a)_n
U(a+n,b+n,x).
\end{align}

Several recurrence relations follow from the confluent hypergeometric equation and its contiguous identities. For Kummer's function, two useful examples are
\be
\left(
x\frac{\dd}{\dd x}+a
\right)
M(a,b,x)
=
aM(a+1,b,x),
\ee
and
\be
\left(
x\frac{\dd}{\dd x}+b-a-x
\right)
M(a,b,x)
=
(b-a)M(a-1,b,x).
\ee
Equivalently,
\[
\begin{aligned}
0={}&
(b-a)M(a-1,b,x)
+
(2a-b+x)M(a,b,x)
\\
&-
aM(a+1,b,x).
\end{aligned}
\]
An identity shifting the second parameter is
\[
b
\left[
M(a,b,x)-M(a-1,b,x)
\right]
=
xM(a,b+1,x).
\]

Useful contiguous relations for Tricomi's function include
\[
U(a,b+1,x)
=
U(a,b,x)
+
aU(a+1,b+1,x),
\]
\[
U(a,b,x)
=
U(a,b-1,x)
+
aU(a+1,b,x),
\]
and
\[
\begin{aligned}
xU(a+1,b+1,x)
={}&
U(a,b,x)
\\
&+
(b-a-1)U(a+1,b,x).
\end{aligned}
\]
Combining the first of these relations with $\dd U(a,b,x)/\dd x=-aU(a+1,b+1,x)$ gives
\[
U(a,b+1,x)
=
\left(
1-\frac{\dd}{\dd x}
\right)
U(a,b,x).
\]
In particular,
\[
U(a,1,x)
=
\left(
1-\frac{\dd}{\dd x}
\right)
U(a,0,x).
\]

We conclude by collecting the identities used most directly in the construction of the mode functions. Define
\[
x=-2iz,
\qquad
a_w=\frac{iw\lambda}{2},
\qquad
w=\pm1.
\]
The derivative identities for the Tricomi branch become
\[
\frac{\dd}{\dd z}
U(a_w,1,-2iz)
=
-w\lambda\,
U(a_w+1,2,-2iz),
\]
and
\[
\left(
\frac{\dd}{\dd z}+2i
\right)
\left[
z\,U(a_w+1,2,-2iz)
\right]
=
-U(a_w,1,-2iz).
\]
For the Kummer branch, the corresponding relations are
\[
\frac{\dd}{\dd z}
M(a_w,1,-2iz)
=
w\lambda\,
M(a_w+1,2,-2iz),
\]
and
\[
\left(
\frac{\dd}{\dd z}+2i
\right)
\left[
z\,M(a_w+1,2,-2iz)
\right]
=
M(a_w,1,-2iz).
\]

With the relative signs adopted in the definitions of $f_X^{(w)}$, $g_X^{(w)}$, and $h_X^{(w)}$, these identities imply
\[
\frac{i}{2}
\frac{\dd}{\dd z}g_X^{(w)}
-
g_X^{(w)}
=
-f_X^{(w)},
\]
and
\[
\frac{\dd}{\dd z}f_X^{(w)}
=
-w\lambda h_X^{(w)},
\]
for both $X=U$ and $X=M$. Since the dressing operator $\widehat{\mathcal D}_\nu^{(w)}$ is a function of $\dd/\dd z$ with constant coefficients, it commutes with $\dd/\dd z$. These identities therefore remain valid after the conformal seeds are dressed by $\widehat{\mathcal D}_\nu^{(w)}$.

\section{Canonical inner product and normalization of the mode basis}
\label{app:inner-product}

In this appendix we introduce the conserved inner product used to normalize the mode basis in Section~\ref{sec:bogoliubov}. Consider the field space vector collecting the mode functions solutions $\mathcal F^{(w)}_X(k,z)$ and $\mathcal S_X^{(w)}(k,z)$ with $X = \{ U , M \}$:
\be
\boldsymbol{\Psi}^{(w)}_X(k,z)
=
\begin{pmatrix}
\mathcal F^{(w)}_X(k,z)\\
\mathcal S^{(w)}_X(k,z)
\end{pmatrix} . \label{app:mode-vector}
\ee
The canonical symplectic product associated with the quadratic action is
\begin{align}
\left(\boldsymbol{\Psi}_X,\boldsymbol{\Psi}_Y\right)
\equiv{}&
i\int \dd^3x\,a^3
\Big[
\mathcal F_X^* D_t\mathcal F_Y
-
(D_t\mathcal F_X)^*\mathcal F_Y
\nn\\
&
\qquad\qquad
+
\mathcal S_X^*\dot{\mathcal S}_Y
-
\dot{\mathcal S}_X^*\mathcal S_Y
\Big].
\label{inner-product-position}
\end{align}
After going to Fourier space and using \(D_t=-HzD_z\), this becomes the conserved Wronskian
\begin{align}
\left(\boldsymbol{\Psi}_X,\boldsymbol{\Psi}_Y\right)
={}&
-i\,\frac{k^3}{H^2z^2}
\Big[
\mathcal F_X^*D_z\mathcal F_Y
-
(D_z\mathcal F_X)^*\mathcal F_Y
\nn\\
&
\qquad\qquad
+
\mathcal S_X^*\partial_z\mathcal S_Y
-
(\partial_z\mathcal S_X)^*\mathcal S_Y
\Big],
\label{inner-product-z}
\end{align}
where $D_z\mathcal F
=
\partial_z\mathcal F
+
\frac{\lambda}{z}\mathcal S$. Then, using the equations of motion, one verifies directly that
\be
\dv{z}\left(\boldsymbol{\Psi}_X,\boldsymbol{\Psi}_Y\right)=0 .
\ee
The reference modes \(u_0\) and \(u_\mu\) are normalized so that their positive-frequency components have unit norm and their complex conjugates have negative norm. Therefore, if a solution is written in terms of the Bogoliubov coefficients as in Eqs.~(\ref{def-ZsX}) and (\ref{def-SsX}):
\begin{align}
\mathcal F_X
&=
f_Xu_0+\bar f_Xu_0^*,
\\
\mathcal S_X
&=
s_Xu_\mu+\bar s_Xu_\mu^*,
\end{align}
then the inner product reduces to:
\be
\left(\boldsymbol{\Psi}_X, \boldsymbol{\Psi}_Y\right)
=
f_X^*f_Y
-
\bar f_X^*\bar f_Y
+
s_X^*s_Y
-
\bar s_X^*\bar s_Y .
\label{inner-product-bogoliubov}
\ee
Moreover, with the normalization chosen in Eqs.~\eqref{f-sol-1} and~\eqref{f-sol-2}, the reconstructed modes entering (\ref{app:mode-vector}) are found to  satisfy
\begin{align}
\left(\boldsymbol{\Psi}_U^{(w)},\boldsymbol{\Psi}_U^{(w')}\right)
&=
\delta_{ww'},
\\
\left(\boldsymbol{\Psi}_M^{(w)},\boldsymbol{\Psi}_M^{(w')}\right)
&=
-\delta_{ww'},
\\
\left(\boldsymbol{\Psi}_U^{(w)},\boldsymbol{\Psi}_M^{(w')}\right)
&=
0 .
\label{UM-orthonormality}
\end{align}
Thus the \(U\)-sector has positive norm, while the \(M\)-sector has negative norm. Expanding the two physical mode doublets as
\be
\boldsymbol{\Phi}_b
=
\sum_{w=\pm1}
\left[
A_b^{(w)}\boldsymbol{\Psi}_U^{(w)}
+
B_b^{(w)}\boldsymbol{\Psi}_M^{(w)}
\right],
\ee
and using the orthonormality relations~\eqref{UM-orthonormality}, the canonical constraints imply the pseudo-unitarity condition
\be
\sum_{b=1}^{2}
\left(
A_b^{(w)}A_b^{(w')*}
-
B_b^{(w)}B_b^{(w')*}
\right)
=
\delta_{ww'}.
\label{app-AA-BB-cond}
\ee
The remaining Bogoliubov constraints similarly give the complementary relation
\be
\sum_{b=1}^{2}
\left(
A_b^{(w)}B_b^{(w')*}
-
B_b^{(w)}A_b^{(w')*}
\right)
=
0 .
\label{app-AB-BA-cond}
\ee

\end{appendix}

\end{document}